\documentclass[conference]{IEEEtran}

\usepackage{ifthen}
\newboolean{is_acm}
\setboolean{is_acm}{false}

\usepackage[english]{babel}
\usepackage{pifont}
\usepackage{makecell}

\usepackage{amsmath}
\usepackage{url}
\usepackage{amssymb,amsfonts}
\usepackage{algorithmic}
\usepackage{graphicx}
\usepackage{textcomp}
\usepackage[dvipsnames]{xcolor}
\usepackage{array}
\usepackage{booktabs}
\usepackage{arydshln}
\usepackage[nopostdot,nonumberlist,style=tree]{glossaries}
\usepackage{tcolorbox}
\usepackage[normalem]{ulem}
\usepackage{pgfplots}
\usepackage{lscape}
\usepackage{capt-of}
\usepackage{etoolbox} 
\usepackage{hyperref} 
\usepackage{cite}
\usepackage[T1]{fontenc}

\usepackage[compatibility=false,font=footnotesize,labelfont=bf]{caption}

\usepackage{tikz}
\usepackage{tikz-qtree}
\usetikzlibrary{shadows, positioning, shapes.multipart, fit}

\usepackage{amsthm}
\newtheoremstyle{boldheadonly}
  {\topsep}    
  {\topsep}    
  {\itshape} 
  {}           
  {\bfseries}  
  {.}          
  { }          
  {}           

\theoremstyle{boldheadonly}

\newtheorem{takeaway}{Takeaway}[section]
\newtheorem{remark}{Remark}[section]

\newcolumntype{C}[1]{>{\centering\arraybackslash}m{#1}}

\pgfplotsset{compat=1.18}

\usepackage{afterpage}
\newcommand{\ra}[1]{\renewcommand{\arraystretch}{#1}}
\newcommand{\good}[0]{\ding{51}}
\newcommand{\bad}[0]{\ding{55}}

\newcommand{\nota}[0]{$-$}
\newcommand{\tablefont}[0]{\footnotesize}

\newcommand{\mr}[1]{\makecell[tl]{#1}}

\usepackage{multirow}
\usepackage{hhline}
\newcommand{\ycell}[2]{
{\multirow{#1}{*}{
  \begin{tabular}{@{}c@{}}%
    \rotatebox[origin=c]{270}{#2}%
  \end{tabular}%
}}}

\newcommand{\yycell}[2]{
{\multirow{#1}{*}{
  \begin{tabular}{@{}c@{}}%
    \rotatebox[origin=c]{90}{#2}%
  \end{tabular}%
}}}
\begin{document}

\title{A Survey of End-to-End Modeling for Distributed DNN Training: Workloads, Simulators, and TCO}

\author{
\IEEEauthorblockN{
Jonas Svedas$^\dagger$, Hannah Watson$^\dagger$, Nathan Laubeuf$^\ddagger$, Diksha Moolchandani$^\ddagger$,\\
Abubakr Nada$^\ddagger$, Arjun Singh$^\dagger$, Dwaipayan Biswas$^\ddagger$, James Myers$^\dagger$, Debjyoti Bhattacharjee$^\ddagger$
}
\IEEEauthorblockA{
\textit{$^\dagger$imec, 20 Station Road, Cambridge CB1 2JD, UK} \\ 
\textit{$^\ddagger$imec, Kapeldreef 75, 3001 Leuven, Belgium} \\ 
firstname.lastname@imec.be}
}
\maketitle
\begin{center}
    \textit{Preprint.}
\end{center}
\begin{abstract}
Distributed deep neural networks (DNNs) have become a cornerstone for scaling machine learning to meet the demands of increasingly complex applications. However, the rapid growth in model complexity far outpaces CMOS technology scaling, making sustainable and efficient system design a critical challenge. Addressing this requires coordinated co-design across software, hardware, and technology layers. Due to the prohibitive cost and complexity of deploying full-scale training systems, simulators play a pivotal role in enabling this design exploration.
This survey reviews the landscape of distributed DNN training simulators, focusing on three major dimensions: workload representation, simulation infrastructure, and models for total cost of ownership (TCO) including carbon emissions. It covers how workloads are abstracted and used in simulation, outlines common workload representation methods, and includes comprehensive comparison tables covering both simulation frameworks and TCO/emissions models, detailing their capabilities, assumptions, and areas of focus. In addition to synthesizing existing tools, the survey highlights emerging trends, common limitations, and open research challenges across the stack. By providing a structured overview, this work supports informed decision-making in the design and evaluation of distributed training systems.
\end{abstract}
\section{Introduction}\label{sec:intro}

\noindent The phenomenal growth in demand for compute power to satisfy ever growing DNN workload requirements~\cite{epoch2024trainingcompute} is leading to rapid growth in data centers and associated energy consumption. Training large language models such as Llama 3, GPT-4 or Gemini Ultra consumes energy costing tens of millions of dollars on state-of-the-art GPU hardware and requires millions of compute hours~\cite{cottier2024risingcosts}. The growing mismatch between compute demand and supply has made training these AI models exceedingly difficult, with significant economic, geopolitical, and societal implications~\cite{yang2025ai}.

Given the slow down in Moore's Law~\cite{hellings23} and the end of Dennard scaling~\cite{esmaeilzadeh2011dark}, the semiconductor community is now engaged in system-technology co-optimisation~(STCO)~\cite{biswas2024stco} at both research and product levels. A better understanding is necessary of algorithms, architecture and mapping techniques to determine wider system-level performance benefits in partnership with new technology developments. While GPUs are the architecture of choice for deep neural network (DNN) training, the landscape is diversifying with the emergence of domain-specific architectures (DSAs) and accelerators. For example, Tensor Processing Units (TPUs)~\cite{jouppi2017datacenter} are widely deployed for both training and inference, while Compute-in-Memory (CiM) accelerators~\cite{wolters2024memory} are being actively researched, primarily for inference and potentially for training in the future~\cite{silvano2023survey, machupalli2022review}. Furthermore, innovations on the non-compute aspects of the system, such as memory technologies~\cite{zacariasmemory} and memory organizations~\cite{wang2022enabling}, as well as interconnects~\cite{jouppi2023tpu}, also have a significant impact at the system level. 

Simulation has become an essential tool in hardware system design, driven by the increasing complexity and cost of modern architectures. This trend aligns with the \textit{shift-left} practice~\cite{shiftleft}, which emphasizes performing design space exploration and performance optimization earlier in the development cycle. By leveraging simulation at early stages, engineers can make informed architectural decisions and evaluate trade-offs before committing to hardware implementation, ultimately reducing engineering time and cost.

Architectural simulators are widely employed to analyze system performance metrics. Typically, a simulator takes a \textit{workload} and a \textit{system configuration} as input and, as output, generates \textit{statistics} (e.g., performance counters) to estimate the behavior of the target system. The fidelity of these statistics depends on the level of detail modeled. Increasing simulation precision often comes at the cost of longer runtime, making the trade-off between speed and fidelity a critical design decision.  However, using traditional architectural simulators~\cite{akram2019survey} for such evaluations is impractical due to their limited scalability and prohibitively slow simulation speeds. At the same time, running experiments on real hardware is often infeasible due to the enormous costs of owning or renting such infrastructure. These challenges have led to the emergence of \emph{distributed DNN training simulators}, which are explicitly designed to estimate system performance and address simulation speed at scale. While simulators are critical for early-stage design space exploration, they inherently involve trade-offs. As noted by Nowatzki et al.~\cite{nowatzki2015architectural}, common challenges include oversimplified abstractions, limited hardware coverage, and difficulties in validation. These trade-offs become even more significant at the scale of modern distributed DNN training.

This survey reviews the emerging class of distributed DNN training simulators. Based on an extensive review of the field, it identifies three key areas that are critical for understanding the design and the future trajectory of these simulators\ifthenelse{\boolean{is_acm}}{ (see Figure~\ref{fig:survey-structure})}{}:
\begin{itemize}
    \item Section~\ref{sec:workload}: \emph{Workload representation}: examines how DNN workloads are constructed and represented for simulation. Given the complexity of modern ML software stacks, understanding workload representation is more important than ever for ensuring simulator efficiency, accuracy, and compatibility with existing ML frameworks.
    
    \item Section~\ref{sec:simulators}: \emph{Distributed DNN training simulators}: presents a taxonomy along with a detailed comparative analysis of existing simulators, highlighting emerging trends, current limitations, and open research gaps.

    \item Section~\ref{sec:tco}: \emph{Total Cost of Ownership (TCO) and environmental cost modeling} highlights the rising importance of TCO and carbon emissions as a figure of merit for assessing the long-term cost efficiency of DNN training systems. Existing TCO and emissions models are reviewed and classified, and the effort necessary to enable their integration into distributed DNN simulation frameworks is discussed.
\end{itemize}

Guo et al.~\cite{guo2024survey} provide a focused survey of distributed DNN training simulators, primarily emphasizing analytical models and a limited set of operator-level intermediate representation (IR) simulators. Their work centers on the mathematical formulations used for performance estimation, offering detailed insight into model-based approaches. 

Complementary to their work this survey evaluates a broader range of simulation frameworks and provides explicit comparisons across hardware-specific compute modeling, network abstraction fidelity, and memory hierarchy modeling. It also bridges the gap between workload representation, performance-oriented simulation, and TCO-aware modeling. Furthermore, it identifies underexplored areas and emerging research opportunities across the stack. By offering a structured overview of distributed DNN training simulators, this survey clarifies key design trade-offs, modeling capabilities, and limitations to support informed tool selection and guide future simulation research.

\ifthenelse{\boolean{is_acm}}{

\begin{figure*}[htbp]
\centering
\includegraphics[width=0.95\linewidth]{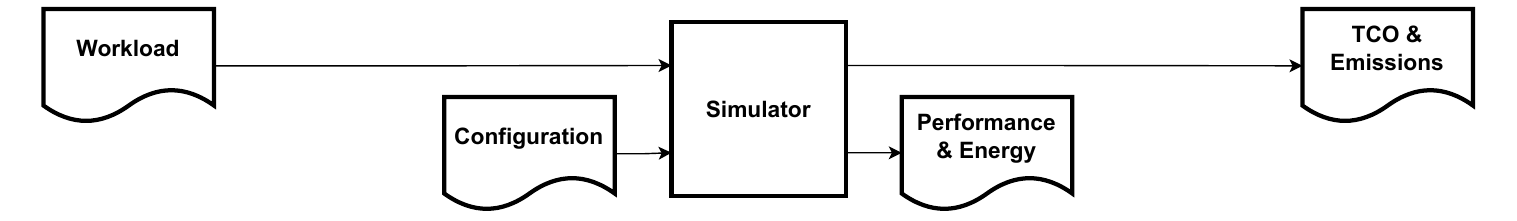}
\begin{tikzpicture}[
  node distance=0.4cm and 0.8cm,
  main/.style={rectangle, draw=black, thick, text width=3cm, align=center, fill=gray!15, font=\small },
  middle/.style={rectangle, draw=black, thick, text width=5cm, align=center, fill=gray!15, font=\small},
  sub/.style={rectangle, draw=black, thin, text width=3cm, align=left, font=\small, fill=white},
  midsub/.style={rectangle, draw=black, thin, text width=5cm, align=left, font=\small, fill=white}
]
\node[main] (sec2) {\textbf{Section}~\ref{sec:workload} \\ DNN Workload Representation for Simulators};
\node[middle, right=of sec2] (sec3) {\textbf{Section}~\ref{sec:simulators} \\ Distributed DNN\\Training Simulators};
\node[main, right=of sec3] (sec4) {\textbf{Section}~\ref{sec:tco} \\ Total Cost of Ownership and Environmental Cost Modeling};
\end{tikzpicture}

\begin{tikzpicture}[remember picture, overlay]
  \draw[black, thick, dotted] ( 3 , 0.25) -- ( 3, 4.4);
  \draw[black, thick, dotted] (-3, 0.25) --  (-3, 4.4);
\end{tikzpicture}

\caption{Overview of the survey structure, organized around the inputs and outputs of an architectural simulator. Section~\ref{sec:workload} focuses on workload representations. Section~\ref{sec:simulators} surveys distributed DNN training simulators. Section~\ref{sec:tco} examines TCO and environmental costs.}
\label{fig:survey-structure}
\end{figure*}

}{
}
\section{DNN Workload Representation for Simulators} \label{sec:workload}

In distributed DNN training simulators, the workload is a critical input, shaping not only the simulation results but also the simulator’s software architecture. A precise understanding of how a DNN workload is represented is essential when designing or using such simulators, as it directly affects simulation accuracy, performance, scalability, and architectural choices.

This section begins by introducing the fundamentals of DNN models and explaining why training is significantly more demanding than inference. It then explores the motivation behind distributing DNN training, the available distribution strategies, and the collective communication algorithms involved. Finally, it categorizes and reviews the workload representations commonly used to simulate distributed DNN training.

\subsection{Background on DNN Training} \label{sec:background}

Deep Neural Networks (DNNs) are commonly formalized as parameterized functions of the form \( y = f(x; \theta) \), where \( x \) is the input data, \( \theta \) denotes the learnable parameters, and \( y \) is the output~\cite{Goodfellow-et-al-2016}. These functions are generally defined as sequence of operations such as matrix multiplications, convolutions, and nonlinear activations, exhibiting a clear dataflow structure. Due to this dataflow structure, the outputs \( y \) of a neural network are computed from its inputs \( x \) through a process known as \emph{forward propagation}. When the forward propagation of a trained network is used for prediction on unseen data, it is often referred to as \emph{inference}.

\begin{figure}[htbp]
    \centering
    \includegraphics[width=0.95\linewidth]{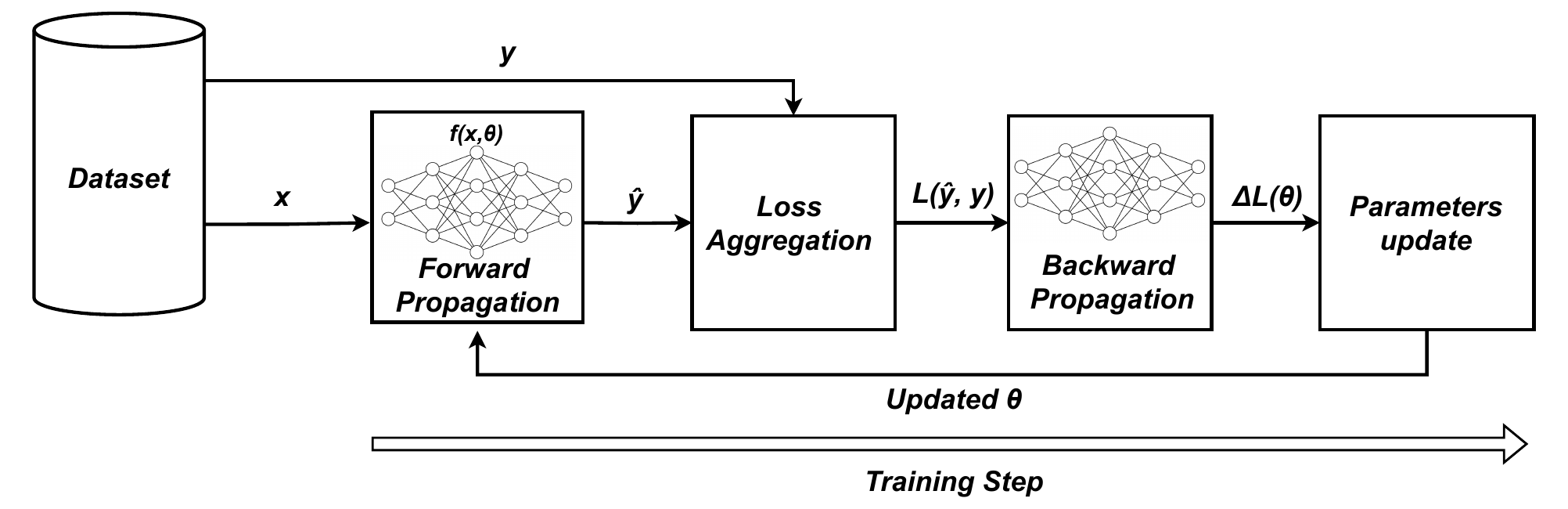}
    \caption{Overview of the DNN training and inference pipeline, showing forward propagation, loss aggregation, backward propagation, and parameters \(\theta\) update.}
    \label{fig:training-step}
\end{figure}

To adapt the parameters \( \theta \) for improved predictive accuracy, \textit{training} is performed, which extends forward propagation with three additional stages: \emph{loss aggregation}, \emph{gradient computation} via \emph{backpropagation}~\cite{rumelhart1986learning}, and \emph{parameter updates}. During training, a batch of input samples is processed in a single \textit{training step}. The model generates predictions for all samples in the batch, and the aggregate prediction error is quantified using a loss function. Gradients of the loss function with respect to the parameters \(\theta\) are then computed by application of the chain rule via backpropagation. The gradients are then used by the update rules of gradient-based optimization algorithms such as Stochastic Gradient Descent (SGD~\cite{ruder2016sgd} or adaptive methods like Adam~\cite{kingma2014adam} to update the model's parameters. These gradients indicate how each parameter should be adjusted to minimize the prediction error, and the parameters are updated accordingly. This sequence of \emph{forward propagation}, \emph{loss aggregation}, \emph{backward propagation}, and \emph{parameters update} completes one \emph{training step} (see Figure~\ref{fig:training-step}). 

Training fundamentally differs from inference by requiring the computation and storage of gradients, along with additional optimizer state variables such as momentum terms~\cite{sutskever2013importance}, adaptive learning rate parameters, and running averages for batch normalization~\cite{ioffe2015batch}. These additional states significantly increase both memory usage and computational complexity during training.

In practice, training modern LLMs requires not one but hundreds of thousands of training steps~\cite{llama405B2024} to converge to an acceptable solution, each involving passes over batches of data. As model sizes and datasets grow, a single training step can become prohibitively slow and memory-intensive. To address this, \textit{parallelism} is employed, enabling training to be \textit{distributed} across multiple compute devices (e.g., GPUs, TPUs)~\cite{dean2012large}. Parallelism introduces an additional layer of complexity, primarily by requiring communication between different compute nodes. It involves a complex software-hardware co-optimization process to balance computation, memory, and communication costs, depending on the chosen parallelism strategy. Moreover, these design choices must be formally captured to accurately represent the distributed training process in a workload model.

\begin{figure}[htbp]
    \centering
    \includegraphics[width=1\linewidth,  trim=20 0 20 0, clip]{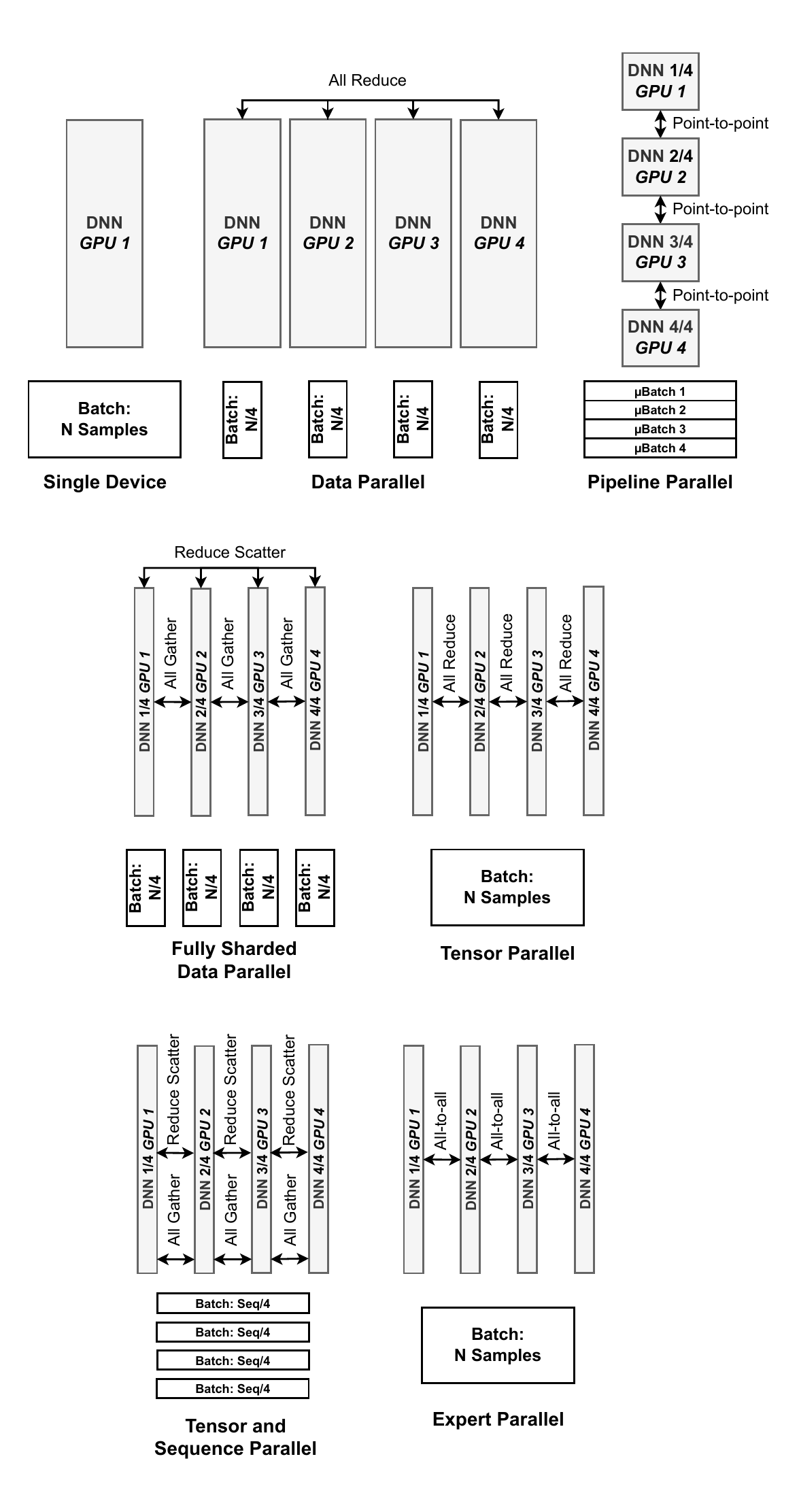}

\caption{
Parallelism strategies in distributed DNN training. 
\textbf{Single Device:} Training runs on a single GPU using the full batch. 
\textbf{Data Parallelism (DP):} Each GPU holds a model replica and processes a unique input shard; gradients are synchronized via \texttt{All-Reduce}. 
\textbf{Fully Sharded Data Parallelism (FSDP):} The model is sharded across GPUs; inputs are split; parameters, activations, optimizer states, and gradients are materialized as needed using \texttt{All-Gather}, with gradients reduced via \texttt{Reduce-Scatter}. 
\textbf{Tensor Parallelism (TP):} Layers are split across GPUs; \texttt{All-Reduce} is used for intermediate results and gradient accumulation. 
\textbf{Tensor + Sequence Parallelism (TP + SP):} Like TP, but inputs are also split along the sequence dimension; uses \texttt{All-Reduce} and \texttt{All-Gather}. 
\textbf{Pipeline Parallelism (PP):} Model layers are distributed across devices; microbatches move through the pipeline via point-to-point communication. 
These strategies can be combined for \emph{hybrid parallelism} (HP) at large scale.
}
\label{fig:parallel}
\end{figure}

Distributed DNN training has employed six principal parallelization strategies: data parallelism~(DP), fully sharded data parallelism~(FSDP), tensor parallelism~(TP), sequence parallelism~(SP), expert parallelism~(EP), and pipeline parallelism~(PP), or hybrid combination of the six \cite{dean2012large, korthikanti2023nvidia_SP, jacobs2023deepspeed_ulysses, singh2023hybrid_TP_EP_DP, rajbhandari2022deepspeed-moe}, as illustrated in Figure~\ref{fig:parallel}. Each method partitions computation across multiple devices, introducing synchronization requirements that are typically handled using communication collectives~\cite{nvidia_nccl}—a set of standardized operations involving multiple devices (see Table~\ref{tab:comm_collectives}). Depending on their implementation and interactions, parallelization strategies will come to rely on different communication collectives. Notably, while DP uses \texttt{All-Reduce} to aggregate gradients across model replicas, FSDP will partition parameters, activations, optimizer states, and gradients across workers, using \texttt{All-Gather} operations to materialize the weights and activations when required, and aggregate gradients using \texttt{Reduce-Scatter} to only update local optimizer states and parameters~\cite{rajbhandari2020zero}. From a workload representation perspective, these operations can be modeled at a high level using analytical latency-bandwidth models, or more precisely using detailed data movement patterns and network topology.

\begin{table}[tbp]
\centering
\caption{Common communication collectives and parallelism strategies they are typically used in.}
\label{tab:comm_collectives}
\ra{1.2}
\tablefont
\begin{tabular}{@{}p{2.4cm}p{1.6cm}p{4.5cm}@{}}
\toprule
\textbf{Collective} & \textbf{Commonly Used In} & \textbf{Role} \\ \midrule
\texttt{All-Reduce} & \mr{DP, TP} & Aggregates gradients (DP), or partial results (TP) across devices~\cite{shoeybi2019megatron}. \\
\texttt{All-Gather} & \mr{FSDP,\\TP + SP} & Reconstructs full tensors from partitioned outputs, e.g., after tensor or sequence splitting~\cite{rajbhandari2020zero, korthikanti2023nvidia_SP}. \\
\texttt{Reduce-Scatter} & \mr{FSDP,\\TP + SP} & Efficiently combines reduction and scattering of gradients or sequence activations. Often paired with \texttt{All-Gather}~\cite{rajbhandari2020zero, korthikanti2023nvidia_SP}. \\
\texttt{Broadcast}      & \mr{DP} & Distributes model parameters from a source device to others during initialization or checkpoint loading. \\
\texttt{Point-to-Point} & \mr{PP} & Transfers activations or gradients between pipeline stages using point-to-point communication~\cite{narayanan2021megatron}. \\
\texttt{All-to-All}     & \mr{EP} & Route tokens to their most fitting parallel expert in Mixture-of-Experts~\cite{artetxe2021moe, singh2023hybrid_TP_EP_DP, rajbhandari2022deepspeed-moe}. \\
\texttt{Barrier}        & \mr{All} & Synchronizes all devices at a consistent execution point. Mostly used for coordination or debugging. \\ \bottomrule
\end{tabular}
\end{table}

When distributing DNN models two fundamental classes of operations arise : \emph{computation} and \emph{communication}. 
\begin{itemize} 
\item Computation encompasses operations executed locally within a compute node, such as tensor contractions, matrix multiplications, or activation functions. 
\item Communication, in contrast, involves data movement across nodes, with performance largely dictated by network latency, bandwidth, and contention rather than local compute capabilities. 
\end{itemize} 
Explicitly separating computation from communication provides a critical abstraction for analyzing performance bottlenecks, guiding system co-design, and informing optimization strategies. This separation is foundational in both production systems and simulators. In production frameworks, distinct libraries are used to target each domain: kernel libraries such as cuDNN~\cite{chetlur2014cudnn} or MIOpen~\cite{khan2019miopen} provide target optimized compute primitives for DNNs, while communication libraries like NCCL~\cite{nvidia_nccl}, Gloo~\cite{pytorch_gloo}, or various MPI implementations~\cite{forum1994mpi} offer collective communication operations. Simulators similarly treat computation and communication as distinct layers, enabling fine-grained performance modeling. Accurately capturing the interaction between these layers is essential for simulating overlapping execution, pipelining, and network congestion—factors that critically influence the scalability and efficiency of distributed DNN training. Recent DNN training simulators leverage various workload representations to model these effects with both fidelity and efficiency.

Distributed DNN training workloads can be represented at different levels of abstraction, ranging from high-level analytical or mathematical models to computational graphs or down to machine-level descriptions of executable instructions. The following sections analyze each abstraction category, surveys common representations, and discusses their implications for the fidelity and efficiency of distributed DNN training simulators.

\subsection{DNN Workload Representations}

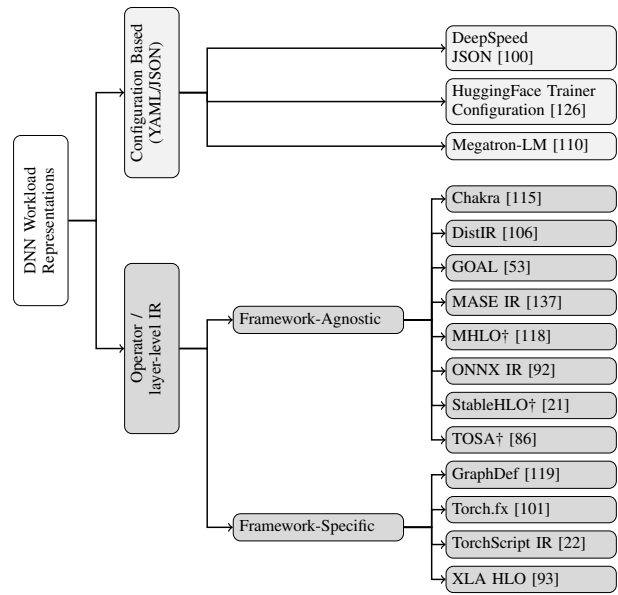
\begin{figure}[htbp]

\label{fig:dnn_repr}
\centering
\scalebox{0.7}{

\begin{tikzpicture}[
    grow'=right,
    level distance=115pt,
    level 1/.style={level distance=60pt},
    level 2/.style={level distance=90pt},
    sibling distance=4pt,    
    every node/.style={draw, rectangle, rounded corners, align=center, font=\small, text width=3cm, anchor=west},
    edge from parent/.style={draw, ->, thick},
    edge from parent path={(\tikzparentnode.east) -- +(15pt,0) |- (\tikzchildnode.west)},
    child anchor=mid,
    cfg/.style= {align=left, draw, rectangle, rounded corners, font=\small, fill=gray!10},
    ir/.style = {align=left, draw, rectangle, rounded corners, font=\small, fill=gray!30},
    ml/.style = {align=left, draw, rectangle, rounded corners, font=\small, fill=gray!60},
]
\Tree
[.\node[text width=0.8cm]{\rotatebox{90}{\parbox{3cm}{\centering DNN Workload\\Representations}}};
    [.\node[text width=0.8cm, cfg]{\rotatebox{90}{\parbox{3cm}{\centering Configuration Based\\(YAML/JSON)}}};
        [.\node[xshift=115, cfg, yshift=81pt]{DeepSpeed JSON~\cite{rasley2020deepspeed}}; ]
        [.\node[xshift=115, cfg]{HuggingFace Trainer\\Configuration~\cite{wolf2020transformers}}; ]
        [.\node[xshift=115, cfg]{Megatron-LM~\cite{shoeybi2019megatron}}; ]
    ]
    [.\node[text width=0.8cm, ir]{\rotatebox{90}{\parbox{3cm}{\centering Operator /\\layer-level IR}}};
        [.\node[ir]{Framework-Agnostic};
            [.\node[ir]{Chakra~\cite{chakra}}; ]
            [.\node[ir]{DistIR~\cite{distir}}; ]
            [.\node[ir]{GOAL~\cite{hoefler2009group}}; ]
            [.\node[ir]{MASE IR~\cite{zhangmase}}; ]
            [.\node[ir]{MHLO\dag~\cite{tensorflow_mlirhlo}}; ]
            [.\node[ir]{ONNX IR~\cite{onnx2019}}; ]
            [.\node[ir]{StableHLO\dag~\cite{openxla_stablehlo}}; ]
            [.\node[ir]{TOSA\dag~\cite{mlir_tosa}}; ]
        ]
        [.\node[ir]{Framework-Specific};
            [.\node[ir]{GraphDef~\cite{tensorflow_graphdef}}; ]
            [.\node[ir]{Torch.fx~\cite{reed2022torchfx}}; ]
            [.\node[ir]{TorchScript IR~\cite{pytorch_torchscript}}; ]
            [.\node[ir]{XLA HLO~\cite{openxla_hlo}}; ]
        ]
    ]
];
\end{tikzpicture}

}

\caption{Representative taxonomy of distributed DNN workload representations, organized into configuration-based, operator / layer-level IRs. \dag denote MLIR~\cite{lattner2021mlir} dialects.}
\label{fig:workloads}
\end{figure}

Workload representations can be broadly classified based on their level of abstraction into two categories: \textbf{configuration-based} and \textbf{operator/layer-level intermediate representations (IRs)}. Figure~\ref{fig:workloads} summarizes currently available workload representations within each category.

\begin{remark}
The objectives of workload representations differ between production environments and simulation. In production, the focus is on achieving high training accuracy, while simulation emphasizes predicting system performance and energy consumption. As a result, the design priorities of workload representations diverge. For instance, IR representations used in simulation may deliberately omit certain details to improve scalability and simulation efficiency.
\end{remark}

\begin{remark}
Workloads can also be represented at a lower machine-level abstraction~\cite{lattner2004llvm, nvidia_ptx, khronos_spirv}. However, none of the reviewed system-level DNN training simulators adopt this approach. This is largely due to simulation speed overhead and limited compatibility with existing ML frameworks, which typically expose computations at the operator or graph level rather than the underlying machine-level backend. Machine-level representations are more commonly employed in node-level architectural simulators focused on microarchitectural analysis~\cite{akram2019survey, jain2018quantitative}.
\end{remark}

\begin{table}[htbp]
\centering
\caption{A summary of intermediate representations with potential for use in DNN simulations. $\dag$ Refers to DistIR simulator that shares the same name with the IR.}
\label{tab:ir_comparison}
\ra{1.2}
\tablefont
\begin{tabular}{@{}lp{1.5cm}p{1.5cm}>{\raggedright\arraybackslash}p{3cm}@{}}
\toprule
\textbf{\makecell[l]{Intermediate\\Representation (IR)}} & \textbf{Portability} & \textbf{\makecell[l]{Parallelism\\Support}} & \textbf{Simulator Integration} \\ \midrule
Chakra~\cite{chakra} & Medium & High & ASTRA-sim~\cite{won2023astrasim2}, LLMServingSim~\cite{llmservingsim} \\ 
DistIR~\cite{distir} & Medium & High & DistIR\dag~\cite{distir} \\ 
GOAL~\cite{hoefler2009group} & Medium & High & ATLAHS~\cite{shen2025atlahs} \\
\makecell[l]{MLIR~\cite{lattner2021mlir}\\$\bullet$ StableHLO~\cite{openxla_stablehlo} \\$\bullet$ MHLO~\cite{tensorflow_mlirhlo}\\$\bullet$ TOSA~\cite{mlir_tosa}} & Medium & Medium & None known\\ 
ONNX~\cite{onnx2019} & High & Low & DistIR\dag~\cite{distir}, ONNXim~\cite{ham2024onnximfastcyclelevelmulticore} (inference only) \\ 
Torch.fx~\cite{reed2022torchfx} & Low & Medium & NeuSight~\cite{lee2024data}, Echo~\cite{feng2024echo} \\ 
TorchScript IR~\cite{pytorch_torchscript} & Low & Low & None known \\ 
XLA HLO~\cite{openxla_stablehlo} & Medium & High & DistIR~\cite{distir} \\ 
\bottomrule
\end{tabular}
\end{table}

\subsubsection{Configuration-Based Representations}
Configuration-based representations describe distributed DNN training setups using high-level metadata files (e.g., YAML, JSON). Common in frameworks like Megatron-LM~\cite{shoeybi2019megatron}, HuggingFace Trainer~\cite{wolf2020transformers}, and DeepSpeed~\cite{rasley2020deepspeed}, they specify model architecture, optimizer settings, parallelism strategies, and hardware placement without detailing tensor-level operations. This format allows users to define hyperparameters, batch sizes, and parallelism modes in a compact, readable form.

For simulation, these representations enable rapid setup and scalability for coarse-grained trend prediction (e.g., time-to-train, memory usage). However, they lack fidelity for fine-grained analysis, as they do not explicitly model details such as operator dependencies, scheduling, and compute-communication overlap.

Configuration-based representations are also relevant in some profiling-based simulators~\ref{fig:simulator_taxonomy}, which profile workloads by deploying ML models through a higher level ML framework (e.g. Megatron-LM, DeepSpeed) and collecting runtime data into a custom, more detailed compute-communication graph. In these cases, abstraction lowering is performed to extract fine-grained details for simulation. However, the underlying scheduling mechanisms in these simulators operate at a more detailed operator level or IR-level granularity; configuration-based representations are primarily used to facilitate trace collection rather than execution.

\subsubsection{Operator/layer-level IR}
Operator- or layer-level IRs offer a detailed and structured representation of DNN workloads by explicitly capturing tensor operations, data dependencies, and data types. Commonly used in domain-specific compilation and serialization pipelines, these IRs bridge high-level model definitions and low-level executable code, enabling both portability and hardware-specific optimizations. This level of abstraction is also well-suited for constructing workload graphs with explicit dependencies, making it effective for both static workload analysis~\cite{ansel2024pytorch2, chen2018tvm, openxla_hlo, rotem2018glow} and simulation.

Operator/layer-level workload representations can be further divided into framework-agnostic and framework-specific categories. A representation is considered {\em framework-agnostic} in the context of this paper, if it can be produced by multiple ML frameworks natively. Framework-agnostic IRs include ONNX~\cite{onnx2019}, MLIR-based dialects such as StableHLO, MHLO, and TOSA, as well as specialized representations like Chakra~\cite{chakra}, DistIR~\cite{distir}, and MASE IR~\cite{zhangmase} as seen in Figure~\ref{fig:workloads}. These formats standardize operators and graph structures to support cross-platform interoperability and optimization. Framework-specific IRs, such as TorchScript IR~\cite{pytorch_torchscript}, Torch.fx graphs~\cite{reed2022torchfx}, and XLA HLO~\cite{openxla_hlo}, are at the core of the tracing, compilation, and serialization features of these frameworks. As such, other IR are often derived, though translation APIs, from these framework-specific IRs.

By explicitly modeling the sequence of operations and communication primitives executed by deep learning frameworks, operator-level IR representations enable more precise simulation of distributed training workloads. They accurately capture various parallelism strategies—such as data parallelism (DP), tensor parallelism (TP), and pipeline parallelism (PP)—along with their synchronization patterns and associated communication overheads. For instance, tensor parallelism in Transformer models typically involves frequent, fine-grained inter-device communication after each matrix multiplication~\cite{shoeybi2019megatron}, which is naturally represented at the IR level. Additionally, operator-level representations facilitate automatic parallelism optimization by enabling efficient analysis and transformations of the workload graph, as demonstrated by systems such as GSPMD~\cite{xu2021gspmd}, PartIR~\cite{alabed2025partir}, FlexFlow~\cite{jia2019flexflow}, and Alpa~\cite{zheng2022alpa}.
Operator-level IR representations strike a good balance between simulation accuracy and scalability, making them widely used in state-of-the-art simulators for distributed DNN training (see Table~\ref{tab:ir_comparison}).

Some workload representations fall between configuration-based and operator-level abstractions. These are typically layer-level IRs, which blend aspects of both approaches. Operating at the granularity of layers rather than individual operators, they encapsulate multiple operations within a single abstraction. Like configuration files, they can be expressed in concise formats such as YAML, JSON, or simple Python scripts, making them easy to read and parse. At the same time, they retain key structural features of operator-level IRs, such as encoding data dependencies, tensor shapes, and high-level data movement patterns. This hybrid design offers a balance between usability and fidelity.

\begin{remark}
MLIR-based representations have emerged as a common foundation across several machine learning compilation stacks, including OpenXLA~\cite{openxla_hlo} and Triton~\cite{tillet2019triton}. While they are not yet adopted by DNN training simulators, their ability to capture model structure and data dependencies across multiple levels of abstraction makes them a compelling option for future use in distributed training simulation. For this reason, they are included in Table~\ref{tab:ir_comparison}.
\end{remark}

\subsection{Takeaways and Future Opportunities}

\begin{takeaway}Extending representations beyond compute/communication to cover infrastructure events like I/O and checkpointing would allow simulators to more realistically model performance slowdowns and energy bottlenecks in real-world training pipelines.
\end{takeaway}

Current workload representations predominantly focus on modeling the core components of distributed training—computation and communication. However, a faithful representation of the full training lifecycle requires additional system-level components such as data loading pipelines, checkpointing mechanisms, fault recovery procedures, and storage system interactions.

Most existing representations either omit or oversimplify these infrastructure aspects. Operator-level approaches, including Chakra~\cite{chakra}, DistIR~\cite{distir}, and GOAL~\cite{hoefler2009group}, offer structured views of compute and communication graphs and present opportunities for extension to system-level modeling. Nonetheless, research to date has prioritized modeling the dominant contributors to training cost—computation and communication—leaving broader infrastructure operations largely underexplored.

Incorporating such operations into workload models would enable simulators to analyze new classes of bottlenecks and trade-offs, such as the effect of checkpointing frequency on throughput or the interaction between I/O bandwidth and data preprocessing latency. Broadening the modeling scope could unlock new insights in the design and optimization of distributed ML systems.

\begin{takeaway}
While numerous standalone operator-level IRs exist (Table~\ref{tab:ir_comparison}), many simulators adopt custom IRs to enable tighter integration and easier implementation of simulator-specific features.
\end{takeaway}

In many profiling-based simulators (Section~\ref{sec:simulators}), the internal representation (IR) is often implemented as a custom directed graph of compute and communication nodes annotated with metadata. Compared to extending an existing IR, this reduces implementation overhead and allows for flexible customization, which is particularly attractive when existing IRs are overly complex or lack simulator-specific capabilities.
However, this flexibility also leads to fragmentation across simulators, as each introduces its own IR. Among these, Chakra stands out as the only framework-agnostic IR that has gained adoption beyond its origin, primarily due to its integration with the widely-used ASTRA-sim.

\begin{takeaway} Torch.fx is a popular IR choice, due to accessibility of Pytorch tracing infrastructure and general availability of ML models written in torch.
\end{takeaway}

While several ML-framework agnostic IRs exist, a common choice in distributed DNN training simulators is torch.fx. This IR is tightly coupled with PyTorch and provides good infrastructure for production model profiling using existing PyTorch tracing infrastructure.

\begin{takeaway}
MLIR is widely used in compiler optimizations, but remains underutilized in simulators, despite its ability to represent computations at multiple levels of abstraction. This presents an opportunity for simulator designers to adopt MLIR-based IRs for modular and scalable workload modeling, especially given their growing use in PyTorch and XLA compilation stacks.
\end{takeaway}

\begin{takeaway}
Machine-level representations are not seen in distributed training simulators likely due to the high overhead of modeling at such fine granularity and the limited relevance of ISA-level details to system-level simulation.
\end{takeaway}

\begin{table}[htbp]
\centering
\caption{Summary of various DNN workload representations.}
\label{tab:workload_comparison}
\tablefont
\resizebox{\linewidth}{!}{
\begin{tabular}{@{}lccc@{}}
\bottomrule
\textbf{Criterion} & \multicolumn{1}{c}{\textbf{\makecell{Configuration-\\Based}}} 
& \multicolumn{1}{c}{\textbf{\makecell{Operator/\\Layer-Level IR}}} 
& \multicolumn{1}{c}{\textbf{\makecell{Machine-\\Level IR}}} \\
\midrule
Fidelity & Low & Medium & High \\ 
Simulator Complexity & Low & Medium & High \\ 
Portability & High & Medium & Low \\ 
Setup Effort & Low & Medium & High \\ 
Simulation Speed & High & Medium & Low \\ 
\toprule
\end{tabular}
}
\end{table}

\section{Distributed DNN Training Simulators} \label{sec:simulators}

Building upon workload representations discussed in Section~\ref{sec:workload}, this section surveys distributed deep neural network (DNN) training simulators--tools that use a workload representation and system configuration as primary inputs to enable performance evaluation of large-scale hardware systems. This section introduces and motivates distributed DNN training simulators, synthesizes existing frameworks, and highlights emerging trends and opportunities in this space.

\subsection{Background and Motivation}
Simulating distributed DNN training is a complex but essential task for advancing the design of machine learning systems. Distributed training spans a wide range of system scales—from single-node, multi-GPU setups to large-scale clusters with thousands of accelerators—and requires tight coordination between computation, communication, and memory subsystems. The design space is vast and tightly coupled across hardware and software layers, making physical prototyping expensive and inflexible. Simulation offers a controllable and cost-effective means to explore this space, but existing tools vary widely in modeling abstraction, fidelity, and integration scope. This section surveys distributed DNN training simulators, organizes them using a taxonomy, evaluates them across common design criteria, and identifies core trade-offs across simulator classes.

The fidelity–speed trade-off is particularly pronounced in distributed DNN training simulators, where even training on real hardware can require days or weeks to complete. The primary challenge lies in the immense scale and complexity of both software and hardware stacks. For instance, the LLaMA~3.1 model (405B parameters) uses a dense transformer architecture trained on 15 trillion multilingual tokens, distributed across 16,384 NVIDIA H100 GPUs—each with 700W TDP and 80GB of HBM3 memory—resulting in a total compute budget of $3.8 \times 10^{25}$ floating-point operations (FLOPs)~\cite{llama405B2024}. Despite major system and software optimizations, training still required 54 days, achieving roughly 90\% effective training time and GPU utilization rates between 38\% and 41\%, depending on the parallelization strategy employed.

Modern machine learning training infrastructures, such as the NVIDIA DGX H200~\cite{nvidiadgxh200}, further highlight this complexity. A single DGX H200 node includes eight H200 GPUs with 1128GB of HBM3e memory, dual CPUs with 56 cores each, 2TB of system RAM, 32GB of cache, and 900GB/s GPU-to-GPU NVLink bandwidth. At this scale, cycle-accurate simulation becomes prohibitively expensive in both time and compute resources. Consequently, distributed DNN training simulators must carefully balance fidelity and performance, selectively abstracting or simplifying components to achieve tractable runtimes while preserving sufficient accuracy for meaningful analysis.

\begin{figure*}[htbp]
    \centering
    \includegraphics[width=0.9\linewidth]{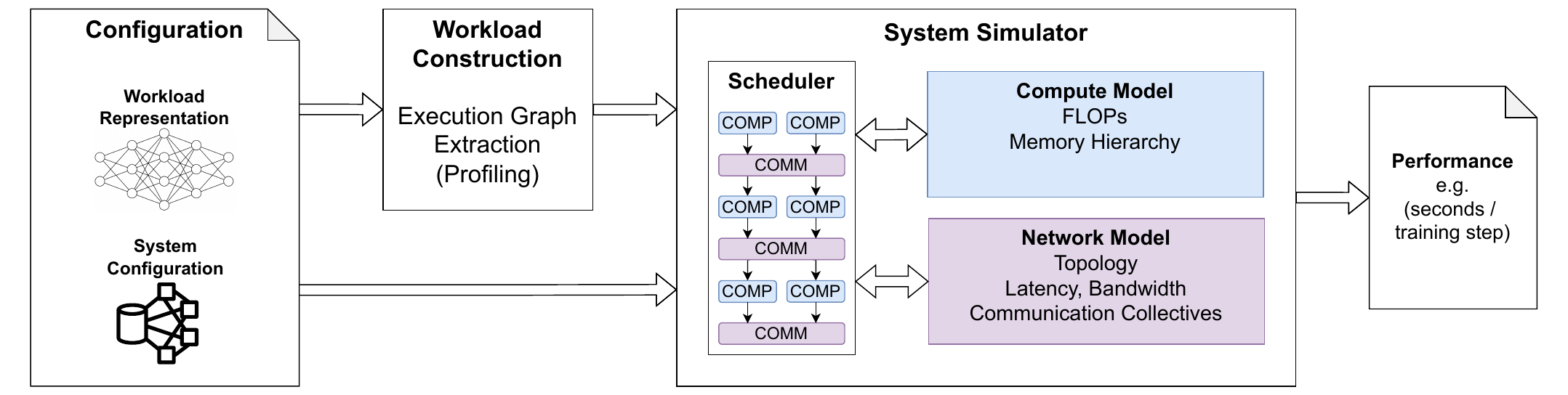}
    \caption{Overview of a typical distributed DNN training simulator architecture.}
    \label{fig:dnn_sim}
\end{figure*}

A typical architecture of a DNN training simulator is illustrated in Figure~\ref{fig:dnn_sim}. The simulation process begins with a user-defined configuration specifying the workload, compute node characteristics, and system topology. In the workload construction phase, an execution graph is generated either through integration with an ML framework or by profiling execution on real hardware. The system simulator models both computation and communication based on this graph, with a scheduler coordinating the interleaving of compute and communication operations. The compute model estimates latency from computational workloads, while the network model captures communication delays between compute nodes. The simulator outputs performance metrics such as training step time.

\subsection{Prior Surveys on Architectural Simulators}

To provide broader context and highlight relevant simulation tools, Table~\ref{tab:arch_sim} summarizes prior surveys of simulators for individual system components. Hwang et al.~\cite{HWANG2025103032} present a comprehensive review of CPU and memory simulators, analyzing architectural models, output metrics, and integration with emerging technologies. Akram et al.~\cite{akram2019survey} focus on simulation frameworks for single- and multi-core CPU systems. For GPUs, Jain et al.~\cite{jain2018quantitative} review simulation methodologies, while Bridges et al.~\cite{bridges2016understanding} examine power modeling and profiling techniques. Patel et al.~\cite{patel2018survey} cover network simulation tools, and additional surveys by Uhlig et al.~\cite{brais2020survey} and Lopez-Novoa et al.~\cite{lopez2014survey} address memory system and accelerator simulation, respectively.

\begin{table}[htbp]
\centering
\caption{Summary of recent surveys on architectural simulators, highlighting their focus areas and key metrics of interest.}
\label{tab:arch_sim}
\ra{1.4}
\tablefont
\begin{tabular}{p{0.7cm}llcccc}
\bottomrule 
\textbf{Year} & \textbf{Survey} & \textbf{Focus Area} &  \multicolumn{4}{c}{\textbf{Scope}} \\ \cmidrule{4-7}
& & &  Perf. & Area & Energy & Acc.$^\dagger$ \\ \midrule
2014 & Lopez-Novoa et al.~\cite{lopez2014survey}      & Accelerator                 &  \good & \good  & \good & \nota \\ 
2016 & Bridges et al.~\cite{bridges2016understanding} & GPU power &  \good & \nota  & \good & \nota \\ 
2018 & Jain et al.~\cite{jain2018quantitative}        & GPU           &  \good & \nota   & \nota  & \good \\ 
2018 & Patel et al.~\cite{patel2018survey}            & Network                       &
\good & \nota & \nota & \nota\\ 
2019 & Akram et al.~\cite{akram2019survey}            & \mr{Single- \&\\multi-core CPU}   &
\good & \nota  & \good & \good \\ 
2020 & Uhlig et al.~\cite{brais2020survey}            & Cache                        & 
\good & \nota & \good & \nota  \\ 
2025 & Hwang et al.~\cite{HWANG2025103032}            & CPU \& mem.               &
\good & \nota & \nota & \nota  \\ 
\toprule
\end{tabular}\\
{ \footnotesize $\dagger$ The accuracy of the simulator output w.r.t the actual hardware output. }

\end{table}

\subsection{DNN Simulator Taxonomy and Comparison}
To systematically categorize the landscape of distributed deep neural network (DNN) training simulators, a taxonomy based on their underlying simulation fidelity is proposed. As depicted in Table~\ref{tab:sim_comparison}, simulators are classified into three primary categories: \emph{analytical}, \emph{profiling-based}, and \emph{execution-based}. This classification reflects fundamental differences in design trade-offs, modeling fidelity, and scalability. 

\begin{table}[htbp]
\caption{Comparison of simulator categories by key aspects.}
\label{tab:sim_comparison}
\centering
\ra{1.4}
\tablefont
\begin{tabular}{@{}llll@{}}
\toprule
\textbf{Aspect} & \textbf{Analytical} & \textbf{Profiling-based} & \textbf{Execution-based} \\ \midrule
\textbf{Workload} & \makecell[l]{Mathematical\\model} & \makecell[l]{Profiling traces\\Operator-level IR} & Machine-level IR \\
\textbf{Speed} & High & Medium & Low \\
\textbf{Fidelity} & Low & Medium & High \\
\textbf{Scalability} & High & Medium & Low \\
\textbf{Strength} & Fast exploration & Fidelity-speed balance & \makecell[l]{Detailed behavior,\\bottleneck analysis} \\
\textbf{Weakness} & Crude assumptions & \makecell[l]{Needs hardware\\for profiling} & \makecell[l]{Impractical speed\\for large workloads} \\
\bottomrule
\end{tabular}
\end{table}

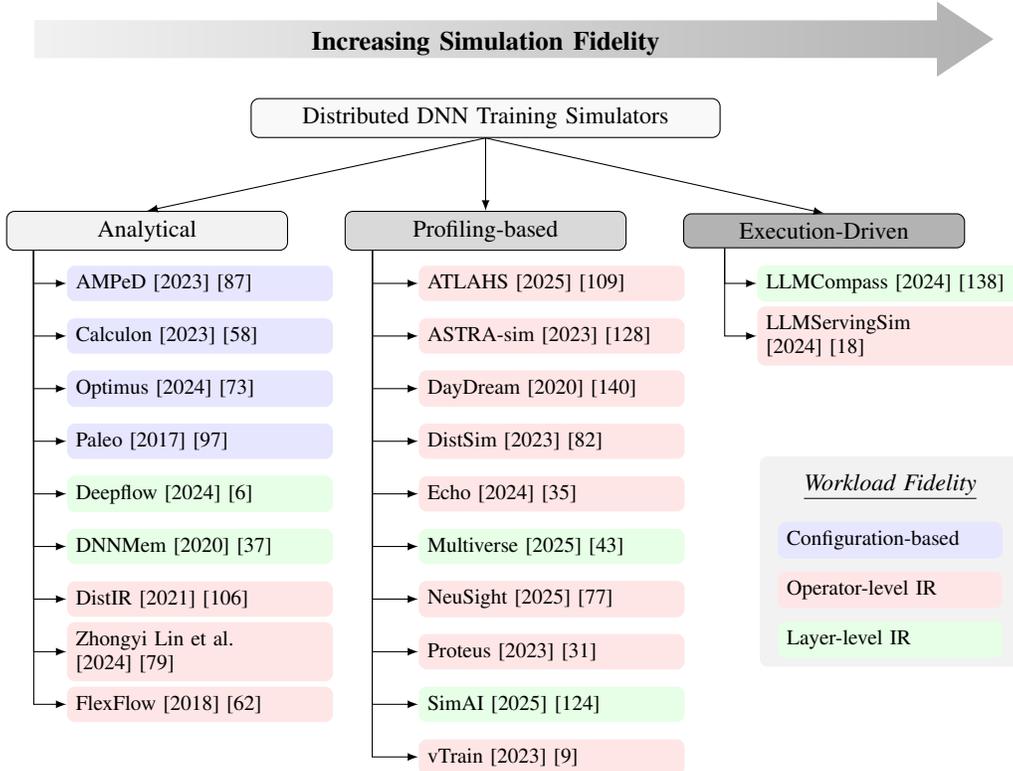
\begin{figure*}[htbp]\
\centering
\begin{tikzpicture}[
  root/.style        = {rounded corners=3pt, thin, text width=3.3cm, font=\small},
  cfg/.style         ={root, font=\footnotesize, fill=blue!10},
  lir/.style         ={root, font=\footnotesize, fill=green!10},
  oir/.style         ={root, font=\footnotesize, fill=red!10},
  analytical/.style  ={align=center, draw, rectangle, fill=gray!10, text width=3.5cm},
  trace/.style       ={align=center, draw, rectangle, fill=gray!30, text width=3.5cm},
  execution/.style   ={align=center, draw, rectangle, fill=gray!60, text width=3.5cm},
  level 1/.style     ={sibling distance=4.5cm},
  edge from parent/.style={->,draw},
  >=latex]

\path[fill=gray!10, left color=gray!10, right color=gray!60]
    (-5,3) -- (7,3) -- (7,3.25) -- (7.75,2.75) -- (7,2.25) -- (7,2.5) -- (-5,2.5) -- cycle;

\node[draw=none, font=\bfseries\normalsize, text=black, inner sep=1pt] at (1,2.7) (nodet) {Increasing Simulation Fidelity};

\node[root,align=center, fill=gray!05, draw, rectangle, text width=6cm, below of=nodet, yshift=0cm]  {Distributed DNN Training Simulators}
  child {node[root, analytical] (a) {Analytical}}
  child {node[root, trace] (t) {Profiling-based}}
  child {node[root, execution] (e) {Execution-Driven}};

\begin{scope}[every node/.style={oir}, node distance=0.7cm]
\node[cfg, below of=a, xshift=20pt] (a1) {AMPeD [2023]~\cite{amped}};
\node[cfg, below of=a1] (a2) {Calculon [2023]~\cite{calculon}};
\node[cfg, below of=a2] (a3) {Optimus [2024]~\cite{optimus}};
\node[cfg, below of=a3] (a4) {Paleo [2017]~\cite{paleo}};
\node[lir, below of=a4] (a5) {Deepflow [2024]~\cite{ardalani2024deepflow}};
\node[lir, below of=a5] (a6) {DNNMem [2020]~\cite{gao2020estimating}};
\node[oir, below of=a6] (a7) {DistIR [2021]~\cite{distir}};
\node[oir, below of=a7] (a8) {Zhongyi Lin et al. [2024]~\cite{lin2024towards}};
\node[oir, below of=a8] (a9) {FlexFlow [2018]~\cite{jia2019flexflow}};
\end{scope}

\begin{scope}[every node/.style={oir}, node distance=0.7cm]
\node[below of=t, xshift=25pt] (t1) {ATLAHS [2025]~\cite{shen2025atlahs}};
\node[below of=t1] (t2) {ASTRA-sim [2023]~\cite{won2023astrasim2}};
\node[below of=t2] (t3) {DayDream [2020]~\cite{zhu2020daydream}};
\node[below of=t3] (t4) {DistSim [2023]~\cite{distsim}};
\node[below of=t4] (t5) {Echo [2024]~\cite{feng2024echo}};
\node[lir, below of=t5] (t6) {Multiverse [2025]~\cite{gui2025accelerating}};
\node[below of=t6] (t7) {NeuSight [2025]~\cite{lee2024data}};
\node[below of=t7] (t8) {Proteus [2023]~\cite{proteus}};
\node[lir, below of=t8] (t9) {SimAI [2025]~\cite{wang2025simai}};
\node[below of=t9] (t10) {vTrain [2023]~\cite{bang2024vtrain}};
\end{scope}

\begin{scope}[every node/.style={oir}, node distance=0.7cm]
\node[lir, below of=e, xshift=25pt] (e1) {LLMCompass [2024]~\cite{llmcompass}};
\node[oir, below of=e1] (e2) {LLMServingSim [2024]~\cite{llmservingsim}};
\end{scope}

\foreach \i in {1,...,9}  \draw[->] (a.190) |- (a\i.west);
\foreach \i in {1,...,10} \draw[->] (t.190) |- (t\i.west);
\foreach \i in {1,...,2}  \draw[->] (e.190) |- (e\i.west);

\matrix [below of=e2, fill=gray!10, yshift=-2cm, rounded corners=3pt, column sep=0.5em, row sep=0.5em, font=\small] {
    \node[draw=none, align=center, text width=3cm, ]{\underline{\em Workload Fidelity}}; \\
    \node[cfg, text width=2.75cm]{Configuration-based}; \\
    \node[oir, text width=2.75cm]{Operator-level IR}; \\
    \node[lir, text width=2.75cm]{Layer-level IR}; \\
};
\end{tikzpicture}
\captionof{figure}{Taxonomy of distributed DNN training simulators categorized into analytical, profiling-based, and execution-driven simulation methodologies.}
\label{fig:simulator_taxonomy}
\end{figure*}

\begin{itemize}
    \item Analytical simulators offer rapid performance estimates with limited precision, relying solely on mathematical models and heuristics. They also exhibit a distinct split based on workload granularity.Simulators such as AMPeD~\cite{amped}, Optimus~\cite{optimus}, and Paleo~\cite{paleo} utilize high-level configurations, resulting in lower modeling fidelity. In contrast, simulators like DistIR~\cite{distir}, Deepflow~\cite{ardalani2024deepflow}, and Calculon~\cite{calculon} employ IR-based inputs, enabling more granular and accurate modeling of computation and communication patterns.

    \item \emph{Profiling-based simulators}, a balance between fidelity and speed by combining empirical data with analytical modeling. Simulators such as ASTRA-sim~\cite{won2023astrasim2}, SimAI~\cite{wang2025simai}, and DistSim~\cite{distsim}, rely on empirical data collected from existing hardware executions to calibrate their models. These simulators often integrate analytical models to extrapolate performance metrics for network, memory, or compute subsystems. This hybrid approach is essential to mitigate the prohibitive costs associated with profiling entire workloads, which would otherwise be comparable to executing full training runs.

    \item \emph{Execution-based simulators} offer high precision through detailed hardware modeling, but at the cost of increased complexity and reduced scalability. Tools such as LLMCompass~\cite{llmcompass} and LLMServingSim~\cite{llmservingsim} include components that capture micro-architectural or instruction-level behavior. However, these detailed models are typically limited to specific subsystems, while the overall simulation relies on higher-level abstractions to balance accuracy and efficiency. This hybrid approach enables focused precision without incurring the prohibitive costs of full-system, cycle-accurate simulation.

\end{itemize}

To further enrich this taxonomy, a secondary dimension is introduced that distinguishes simulators based on the granularity of workload representations they support. Specifically, simulators are differentiated by their use of either high-level configuration inputs or more detailed intermediate representations (IRs). This distinction is visually represented through color-coding in the taxonomy diagram in Figure~\ref{fig:simulator_taxonomy}. Notably, there exists a natural correlation between the fidelity of simulators and the granularity of their input workloads. Higher-fidelity simulators necessitate more detailed workload representations to accurately capture the intricacies of distributed DNN training.

Building upon the proposed taxonomy, a comprehensive comparison table (Table~\ref{tab:dnnsim}) is presented that synthesizes key attributes of existing distributed DNN training simulators. Although some simulators, such as LLMCompass~\cite{llmcompass} and LLMServingSim~\cite{llmservingsim}, focus on inference rather than training, they are included in this survey because of their unique architectural strengths and the potential for future extension to support training workloads. The structured comparison aims to support researchers and practitioners in selecting appropriate simulation tools tailored to their specific requirements, while also helping identify gaps and opportunities for future research and development.

Table~\ref{tab:dnnsim} organizes columns into five main groups. The first group, \textit{Simulator Information}, captures core metadata, including simulator name, year of publication, licensing, and programming language. The next three groups—\textit{Workload}, \textit{Compute Node}, and \textit{Network}—reflect the fundamental pillars around which simulators are typically designed. As the workload fundamentally drives simulator behavior, many tools are structured around the core DNN operators, mapping their execution to compute and communication components. Consequently, most simulators targeting system-level performance modeling feature explicit compute and network modules, alongside a scheduling mechanism that orchestrates operations across these domains. Some simulators additionally model remote memory hierarchies or disaggregated memory systems, motivating the inclusion of a separate \textit{Remote Memory} category in the comparison. The final group, \textit{Total Cost of Ownership} (TCO), reflects an emerging but increasingly important dimension in simulator design. While TCO modeling is still uncommon in architectural simulators, it is a critical area given the growing interest in the economic and environmental sustainability of large-scale DNN training systems.

\afterpage{
\begin{landscape}
\setlength{\abovecaptionskip}{4pt}
\begin{table}[htbp]
\scriptsize
\caption{Summary of distributed DNN training simulators and their main characteristics.}
\label{tab:dnnsim}
\ra{1.2}
\addtolength{\tabcolsep}{-0.2em}
\scalebox{0.95}{
\begin{tabular}{@{}!{\vline}p{0.3cm}!{\vline}llp{1cm}ll!{\vline}lll!{\vline}lll!{\vline}llc!{\vline}c>{\raggedright\arraybackslash}p{1.3cm}@{}!{\vline}}
\hhline{-|-----|---|---|---|--}
\ycell{3}{\textbf{Year}} & \multicolumn{5}{c!{\vline}}{\textbf{SIMULATOR INFO}} & \multicolumn{3}{c!{\vline}}{\textbf{WORKLOAD}} & \multicolumn{3}{c!{\vline}}{\textbf{COMPUTE NODE}} & \multicolumn{3}{c!{\vline}}{\textbf{NETWORK}} & \multicolumn{2}{c!{\vline}}{\textbf{OTHER}} \\ 
\hhline{~|-----|---|---|---|--}
 & \textbf{Name} & \textbf{License} & \textbf{Lang.} & \textbf{\mr{Tool\\Dependency}} & \textbf{Error} & 
\textbf{\mr{Input\\Format}} & \textbf{\mr{Target\\Models}} & \textbf{\mr{Demonstrated\\Models}} &
\textbf{Fidelity} & \textbf{Target HW} & \textbf{Validation} & 
\textbf{Net Model} & \textbf{\mr{Topology\\model}} & \textbf{\mr{Max\\Scalability$^*$}} &
\textbf{\mr{Remote\\Mem.}} & \textbf{\mr{TCO\\Aspects}} \\ 
\hhline{-|-----|---|---|---|--}
\ycell{2}{2017} & \href{https://github.com/TalwalkarLab/paleo}{Paleo}~\cite{paleo} & Apache 2.0 & Python   & \nota & 0.10 & \mr{Layer-\\level IR}  & \mr{CNN,\\GAN} & 
\mr{AlexNet,\\ VGG-16, NiN} & \mr{Layer-wise\\analytical} & Agnostic & \mr{128xK20X,\\8xK20} & Analytical & \nota & 128 & \good & \nota \\ \hdashline[1pt/1pt]

\ycell{1}{2019} & \href{https://github.com/flexflow/flexflow-train}{FlexFlow}~\cite{jia2019flexflow} & Apache 2.0 &  \makecell[l]{C++,\\Python} & \nota   & 0.3$^\ddag$ & Op-level IR  & Agnostic & \mr{AlexNet, Inception-v3,\\ ResNet-101, RNNLM, NMT} & \mr{Analytical} & GPUs & \mr{4xP100,\\ 64xK80} & Analytical & \makecell[l]{any,\\point-to-point\\comms. only} & 64 & \bad & \nota \\
\hdashline[1pt/1pt]
\ycell{3}{2020} & \href{https://github.com/matrix72c/Daydream}{DayDream}~\cite{zhu2020daydream}  & \nota & \makecell[l]{C++,\\Python} & \mr{CUPTI+Caffe,\\MXNet,PyTorch} & 0.10 & \mr{Kernel-\\level IR}  & Agnostic & \mr{BERT, GNMT, Resnet-50} & \mr{Kernel\\trace-driven} & \mr{NVIDIA\\GPU} & 4x2080Ti & Analytical & \nota & 8 & \good & \nota \\
& DNNMem~\cite{gao2020estimating}  & \makecell[l]{Closed\\source} & \nota & \mr{MXNet,PyTorch,\\TensorFlow}& 0.118 & \mr{Layer-\\level IR}  & Agnostic & \mr{BERT, LSTM, VGG16,\\Resnet50, InceptionV3} & 
\mr{Op.-level\\analytical} & \mr{Memory\\only} & 12xK80 & \nota & \nota & 12 & \nota & \nota \\
\hdashline[1pt/1pt]
\ycell{2}{2021} & \href{https://github.com/microsoft/dist-ir}{DistIR}~\cite{distir}  & MIT & Python & \mr{ONNX,\\XLA,HLO} & 0.071$^\dag$ & Op-level IR &  Agnostic & MLP & Analytical & Agnostic & 4xTitan V & Analytical & \nota & 32 & \bad & \nota \\
\hdashline[1pt/1pt]
& \href{https://github.com/astra-sim/astra-sim}{ASTRA-sim}~\cite{won2023astrasim2}  & MIT & C++ & \mr{PyTorch,\\FlexFlow,Chakra,\\ASTRA-sim ET} & 0.05 & \mr{Op-level IR}  & Agnostic & DLRM, GPT-3 & \mr{Analytical,\\Profiling-\\based} & GPU, TPU & 16xV100 & \mr{Analytical\\(incl. congestion),\\NS3, Garnet} & \mr{Multi-level\\switch,\\ring, mesh} & 8192 & \good & \nota \\
\ycell{2}{2023} & \href{https://github.com/CSA-infra/AMPeD}{AMPeD}~\cite{amped}  & MIT & Python & \nota & 0.12$\ddag$ & cfg-based  & LLM & GPT variants & Analytical & GPU & \mr{16xV100} & Analytical & ring & 3072 & \bad & \nota \\
& \href{https://github.com/mikulatomas/distsim}{DistSim}~\cite{distsim}  & MIT & Python & \mr{PyTorch,\\CUPTI} & 0.04 & Op-level IR  & Agnostic & BERT, GPT-2, T5 & \mr{Profiling-\\based} & GPU & 16xA40 & \mr{Profiling-based,\\analytical for >8} & \nota & 128 & \bad & \nota \\
& \href{https://github.com/calculon-ai/calculon}{Calculon}~\cite{calculon}  & Apache 2.0 & Python &\mr{Megatron-\\LLM} & 0.0365 & cfg-based   & \mr{Megatron-\\LLMs} & 
\mr{GPT3, Megatron-1T,\\Turing-NLG} & Analytical & GPU & \mr{A100 (Selene)} & Analytical & \nota & 8192 & \good & CapEx\\
\hdashline[1pt/1pt]
& \href{https://github.com/nanocad-lab/DeepFlow}{Deepflow}~\cite{ardalani2024deepflow}  & Apache 2.0 & \makecell[l]{C,\\Python} & \nota & 0.10 & \mr{Layer-\\level IR}  & Agnostic & 2-layer LSTM & Analytical & GPU, TPU & P4, DGX-1 & Analytical & \mr{Mesh,Torus,\\crossbar} & 512 & \bad & Technology scaling, energy \\
& \mr{\href{https://github.com/PrincetonUniversity/LLMCompass}{LLM Compass}~\cite{llmcompass}}  & BSD-3 & Python & SCALE-Sim & 0.11 & \mr{Layer-\\level IR} &  \mr{LLM\\inference} & GPT-3 & \mr{Execution\\driven} & \mr{A100,\\MI210} & \mr{4xA100,MI210\\8xTPUv3} & \mr{Profiling-\\based} & Analytical & 4 & \bad & {Logic die \& memory cost} \\
& \href{https://github.com/taehokim20/LLMem}{LLMem}~\cite{kim2024llmem}  & \nota & Python & \mr{PyTorch,\\Colossal-AI} & 0.03 & \mr{Layer-\\level IR} &  \mr{LLM\\fine-tuning} & \mr{OPT, BLOOM, CodeGen,\\GPT-neo,BioGPT,\\gpt\_bigcode, Llama} & \mr{Op.-level\\analytical} & GPU & V100 & \nota & \nota & 4 & \nota & \nota \\
\ycell{5}{2024} & \href{https://github.com/JF-D/Proteus}{Proteus}~\cite{proteus}  & \nota & Python & PyTorch & 0.03 & Op-level IR &  Agnostic & \mr{ResNet50, Inception\_V3,\\VGG19, GPT-2,\\GPT-1.5B, DLRM} & \mr{Profiling-\\based} & \mr{NVIDIA\\GPU} & \mr{8x(A100,\\TitanXp, V100)} & Analytical &  \mr{Topology \&\\congestion-\\aware analytical} & 32 & \bad & \nota \\
& \mr{\href{https://github.com/casys-kaist/LLMServingSim}{LLMServing}\\Sim~\cite{llmservingsim}}  & MIT & Python & \mr{ASTRA-sim,\\GeneSys,ONNX\\PolyMath} & 0.15 & Op-level IR   & Inference  & GPT3, Llama & Analytical & NPU & $\oplus$& \mr{Congestion-aware\\analytical,\\NS3, Garnet} & \mr{Mesh, ring, \\switch} & 2048 & \good & \nota \\
& Optimus~\cite{optimus}  & \makecell[l]{Closed\\source} & Python & \nota  & 0.13 & cfg-based &  \mr{LLM\\training} & \mr{GPT, Llama2 variants} & Analytical & GPU & A100, H100 & Analytical & \nota & 4096 & \bad & {Tech. \& mem. scaling} \\
& \href{https://github.com/VIA-Research/vTrain}{vtrain}~\cite{bang2024vtrain}  & MIT & Python & \mr{PyTorch,\\CUPTI} & 0.14 & Op-level IR  & Agnostic & MT-NLG & \mr{Profiling-\\based} & A100 & 8xA100 & Analytical & \nota & 3360 & \bad & \nota \\
& \mr{\href{https://github.com/owensgroup/ml_perf_model}{Lin et al.}~\cite{lin2024towards}} &  BSD-3 & Python & \mr{PyTorch,\\FBGEMM} & 0.0521 & Op-level IR  & Agnostic & \mr{DLRM, BERT,\\GPT-2, XLNet} & Analytical & GPU & \mr{4xA100,\\4xGV100} & Analytical & \nota & 4 & \bad & \nota \\
& Echo~\cite{feng2024echo}  & $\bigtriangleup$ &  \makecell[l]{C++,\\Python} & \mr{PyTorch,\\NCCL} & 0.0275 & Op-level IR &  Agnostic & \mr{BERT\_L, GPT-2,\\VGG19, ResNet152} & \mr{Profiling-\\based} & \mr{NVIDIA\\GPU} & 96xH800 & Analytical & Tree, ring & 8192 & \bad & \nota \\
\hdashline[1pt/1pt]
& \href{https://github.com/spcl/atlahs}{ATLAHS}~\cite{shen2025atlahs} & $\bigtriangleup$ & \nota & \mr{LogGOPSim,\\htsim} & 0.05$\ddag$ & Task-level IR & agnostic & \mr{DLRM, Llama, MoE,\\HPCG, LULESH, LAMMPS,\\ICON, OpenMX} & \mr{Profiling-\\based} & \mr{GPU\\CPU} & \mr{64 x Intel Xeon E5,\\ASTRA-sim} & htsim, NS-3 & \mr{Fat-Tree,\\dragonfly} & 64 &\mr{ storage\\ traffic} & \nota \\
\ycell{5}{2025} & \href{https://github.com/sitar-lab/NeuSight}{NeuSight}~\cite{lee2024data}  & MIT & Python & \mr{PyTorch\\TensorFlow} & 0.077 & Op-level IR &  Agnostic & 
\mr{BERT, GPT2, GPT3,\\OPT, SwitchTrans} & Analytical & Agnostic & \mr{P4, P100, V100,\\T4, A100, H100,\\ L4, AMD MI100–250} & Analytical & \nota & 3840 & \bad & \nota \\
& \href{https://github.com/NASP-THU/multiverse}{Multiverse}~\cite{gui2025accelerating}  & MIT & \makecell[l]{C++,\\Python} & Madrona & 0.03$\ddag$ & \mr{Layer-\\level IR} &  {\small $\diamond$} & GPT-3, Llama & \mr{Profiling-\\based} & \mr{NVIDIA\\GPU} & 1024xH100 & NS-3 based & any & 54000 & \bad & \nota \\
& \href{https://github.com/aliyun/SimAI}{SimAI}~\cite{wang2025simai}  & Apache 2.0 & C++& NS-3, NCCL & 0.019 & Layer-level IR & Agnostic & \mr{GPT-3, Llama3} & \mr{Profiling-\\based} & \mr{NVIDIA\\GPU} & \mr{1024xA100,\\1024xH100} & NS-3 based & any & 1024 & \bad & \nota \\
\toprule
\end{tabular}
} 
\\
\parbox{18cm}{\tiny 
$^\dag$~Computed as $(1 - \text{Pearson correlation})$; not a direct error metric. 
$\bigtriangleup$~Open source release planned. 
$\diamond$~Not described. 
$\oplus$~Relies on ASTRA-sim, GeneSys. 
$\ddag$~Error values reported as maximum error rather than average error. 
$^*$~The maximum simulation scale demonstrated (some simulators such as analytical ones could potentially scale much further).
}

\end{table}
\end{landscape}
} 

\subsection{Takeaways and Future Opportunities}

\begin{takeaway}
    Recent simulators are driven by LLM workloads, but only a subset are designed with the flexibility needed to generalize to other ML architectures.
\end{takeaway}
The recent surge in distributed DNN training simulators is closely linked to the rise of large language models (LLMs), which have introduced significant system complexity and driven demand for scalable performance modeling. Nearly all surveyed tools have emerged post-2020, reflecting this shift. While analytical simulators such as Paleo~\cite{paleo}, AMPeD~\cite{amped}, and Optimus~\cite{optimus} operate at a high level of abstraction, many others—including ASTRA-sim~\cite{won2023astrasim2}, DistIR~\cite{distir}, vTrain~\cite{bang2024vtrain}, and NeuSight~\cite{lee2024data}—rely on ML frameworks for workload representation via intermediate representations (IRs), which are essential for capturing large and heterogeneous operator graphs. Only a few simulators, such as ATLAHS~\cite{shen2025atlahs}, extend support beyond LLMs and CNNs to a broader class of AI and HPC workloads. This flexibility is increasingly valuable for modeling emerging DNN architectures and diverse computational patterns.

\begin{takeaway}
Configuration-based workload representations are becoming less popular over time in distributed DNN simulators.
\end{takeaway}

While they provide a convenient and scalable way to describe high-level operations and parallelism strategies, they do not capture kernel-level execution details or operator dependencies. As simulators increasingly aim to model fine-grained behaviors—such as scheduling, communication overlap, and hardware-specific optimizations—more detailed operator-level representations are becoming necessary for improved accuracy and fidelity.

\begin{takeaway}
Simulators tend to be centered around a single compute target, with NVIDIA GPUs dominating, highlighting a gap in broader hardware generalizability.
\end{takeaway}
While a diverse range of target hardware is supported across simulators, most are effectively centered around a single dominant architecture. Analytical tools such as Paleo~\cite{paleo} and Optimus~\cite{optimus} tend to be more hardware-agnostic due to their higher level of abstraction, enabling flexibility across a broad range of devices, however more detailed simulators especially profiling-based are often tightly coupled to specific machine learning frameworks—such as PyTorch which in turn limits backend support. This is evident in tools like Proteus~\cite{proteus}, Echo~\cite{feng2024echo} and ASTRA-sim ~\cite{won2023astrasim2} where workload construction and trace collection pipelines are narrowly tailored to NVIDIA-based environments. Similarly, simulators like SimAI~\cite{wang2025simai} and MultiVerse~\cite{gui2025accelerating} rely heavily on NVIDIA libraries such as NCCL for modeling network communication. While this ecosystem dependence is understandable given the widespread adoption of NVIDIA GPUs and mainstream ML frameworks, the lack of generalizability across heterogeneous compute architectures remains a key limitation, constraining the evaluation of emerging ML workloads and novel hardware platforms.

\begin{takeaway}
Validation, especially at scale, remains a key weakness.
\end{takeaway}

A persistent challenge across the simulator landscape is validation. Evaluating performance at the scale of thousands of nodes is typically infeasible due to hardware constraints and cost, leading most simulators to rely on indirect validation against published system-level results~\cite{grattafiori2024llama3herdmodels, deepseekai2025deepseekr1incentivizingreasoningcapability}, or to validate small-scale node behavior and assume correctness when scaled up. SimAI~\cite{wang2025simai} and MultiVerse~\cite{gui2025accelerating} stand out as having validated their results on 1024-node A100 GPU clusters. Nevertheless, validation remains a consistent bottleneck across both compute and network domains, limiting confidence in simulator accuracy at scale.

\begin{takeaway}
Most simulators use analytical network models, with few accounting for congestion, topology, or achieving large-scale validation due to data and complexity constraints.
\end{takeaway}
The network modeling component is fundamental to distributed training. The majority of simulators adopt relatively simple analytical models for communication latency, often inspired by the classic \(\alpha + \beta n\) cost model~\cite{culler1993logp}, where \(\alpha\) represents fixed communication latency due to serialization overhead and \(\beta\) represents the inverse of bandwidth (i.e., the per-byte transmission latency due to link delay). These models typically rely on parameterizing \emph{latency}, \emph{bandwidth}, and \emph{hop count}, using either custom estimates or empirical data from communication libraries such as NCCL~\cite{nvidia_nccl}.
While this approach enables scalability, it often overlooks important factors such as network congestion and topology effects, which can significantly impact performance at scale. Only a handful of simulators such as ASTRA-sim~\cite{won2023astrasim2}, Deepflow~\cite{ardalani2024deepflow}, SimAI~\cite{wang2025simai}—incorporate more detailed network modeling that accounts for these effects. Notably, profiling-based approaches to network modeling are rare due to the prohibitive cost and complexity of collecting profiling data at large cluster scales. Among the few exceptions are DistSim~\cite{distsim} and LLMCompass~\cite{llmcompass}, which attempt to model network behavior using data collected from NCCL benchmarks and profilers. SimAI~\cite{wang2025simai} and Multiverse~\cite{gui2025accelerating} are among the first simulators to demonstrate detailed network simulation and validation at very large scale. SimAI introduces a parallelization technique for NS-3, while Multiverse builds on this by restructuring the network simulator using data-oriented design to further improve simulation performance at scale. Network validation remains a major weakness across the field. In general, large-scale experimental data is difficult to obtain due to limited and costly resources, while small-scale benchmarks often fail to capture emergent behaviors that arise only at scale.

\begin{takeaway}
ASTRA-sim~\cite{won2023astrasim2} has emerged as an influential framework for distributed DNN training simulation.
\end{takeaway}

ASTRA-sim’s modular design and built-in API support allow its core scheduler to be extended with external compute, network, and memory models. This flexibility supports multiple levels of abstraction within a single framework, enabling users to combine different compute, memory, and network backends. While other simulators can also be extended, ASTRA-sim’s architecture explicitly prioritizes modularity, which has contributed to its adoption across several recent efforts. Notably, it has been extended in follow-up works such as LLMServingSim~\cite{llmservingsim}, SimAI~\cite{wang2025simai}, and Multiverse~\cite{gui2025accelerating}.

\begin{takeaway}
Python is the dominant language of choice for implementation of DNN simulators due to ML framework integration, while C++ is used when detailed network or compute modeling is needed.
\end{takeaway}

A common design choice in DNN simulators is the use of Python, contrasting with traditional architecture simulators that heavily rely on C++ or SystemC. Python enables tight integration with ML frameworks like PyTorch and TensorFlow, allowing simulators to reuse native intermediate representations and reduce the complexity of modeling large operator graphs. However, simulators such as ASTRA-sim~\cite{won2023astrasim2} and SimAI~\cite{wang2025simai} adopt C++ to support modular integration with detailed compute and network backends like Garnet~\cite{garnet} and NS3~\cite{ns3}. 
While Python accelerates development and prototyping, C++ offers better extensibility and performance for multi-fidelity simulations.

\begin{takeaway}
Despite the relevance of operating energy to large-scale training, this remains a key gap in current simulators.
\end{takeaway}

While some simulators incorporate capital expenditures as a design optimization parameter, none currently model operating energy consumption directly. Among the surveyed tools, vTrain~\cite{bang2024vtrain} is the only simulator that accounts for operating costs indirectly, using cloud service provider pricing (e.g., AWS) as a proxy for estimating the total dollar cost of training—thereby capturing both hardware and energy costs in an approximate manner. This highlights a significant gap in current research, particularly given the substantial energy and environmental impact of large-scale training~\cite{cottier2024risingcosts,schneider2025life}.

Profiling-based simulators such as ASTRA-sim~\cite{won2023astrasim2}, DistSim~\cite{distsim}, and vTrain~\cite{bang2024vtrain} are well-positioned to extend their profiling pipelines for system-level energy estimation, as the collected execution traces could be paired with power models or measurement tools. At the component level, tools like AccelWattch~\cite{kandiah2021accelwattch} estimate GPU power with cycle-level detail, while analytical models such as IPP~\cite{hong2010integrated}, Guerreiro et al.\cite{guerreiro2018gpgpu}, and GPUJoule\cite{arunkumar2019understanding} provide average GPU power estimates. For CPUs, well-established power simulators include McPAT~\cite{li2009mcpat}, Wattch~\cite{brooks2000wattch}, and SimplePower~\cite{ye2000design}. While these tools are not yet integrated into distributed DNN simulators, they represent valuable building blocks for expanding TCO modeling in future research.

\begin{takeaway}
Capital costs are rarely included in the design optimization loop, representing a gap in current simulator capabilities.
\end{takeaway}

Only a small subset of simulators attempt to quantify the capital cost of the simulated hardware. Most assume a fixed hardware configuration and focus primarily on optimizing the training plan or predicting model performance, rather than optimizing the underlying system itself. As a result, capital cost modeling is often deprioritized.

Calculon~\cite{calculon} incorporates basic system-level cost estimates by using off-the-shelf GPU and HBM prices to guide hardware configuration decisions. Deepflow~\cite{ardalani2024deepflow} explores DNN training from the perspective of hardware technology scaling, which—while not a cost model—offers insight into manufacturing trends and their cost implications. LLMCompass~\cite{llmcompass} goes further by modeling detailed manufacturing costs for logic and memory components.

\begin{takeaway}
Local and disaggregated memory remain overlooked targets for simulation research.
\end{takeaway}

Both local and disaggregated (remote) memory are rarely modeled explicitly in current simulators. All simulators, except execution-driven ones~\cite{llmcompass, llmservingsim}, model local memory access only implicitly—typically by embedding memory behavior within operator performance models. This abstraction limits the ability to explore local memory microarchitectures and their impact on training performance.

Two simulators stand out in this space: DNNMem~\cite{gao2020estimating} and LLMem~\cite{kim2024llmem}, both of which focus on estimating the maximum memory requirements of distributed training workloads. However, their goal is capacity estimation rather than performance modeling.

Some simulators—most notably ASTRA-sim 2.0~\cite{won2023astrasim2} and Calculon~\cite{calculon}—explicitly model disaggregated memory and analyze its impact on data movement, bandwidth, and overall training performance. However, comprehensive treatment of memory behavior across both local and remote hierarchies remains limited and underexplored.
\section{Total Cost of Ownership and Environmental Cost Modeling } \label{sec:tco}

As distributed DNN training grows in scale, \emph{Total Cost of Ownership} (TCO) and \emph{environmental cost} have emerged as key concerns alongside traditional metrics like performance and scalability. The infrastructure required to train large language models represents a major investment, making it essential to account for costs over the entire \emph{system lifecycle}. Yet, most existing distributed DNN simulators emphasize performance metrics such as execution time and resource usage, with little attention to economic or environmental impact. Incorporating TCO and environmental cost modeling into these simulators could enable more realistic and actionable evaluations, guiding design choices that balance efficiency, cost, and sustainability.

The remainder of this section, first introduces key TCO concepts and definitions, followed by a brief taxonomy of existing modeling approaches.
It then highlights insights from representative TCO models and discusses how these frameworks can complement DNN simulators, enabling cost- and energy-aware evaluations in future research.

\subsection{Total Cost of Ownership} 
The cost of distributed DNN training systems can rise quickly due to the adoption of advanced hardware and growing system complexity. This poses a significant barrier to accessing state-of-the-art DNN models, especially for smaller organizations and research groups. Accurately quantifying the relationship between system performance and TCO is essential to determine whether investments in new hardware or infrastructure are justified.

\begin{figure*}[htbp]
    \centering
    \includegraphics[width=0.75\linewidth,trim=0 0 0 5,clip]{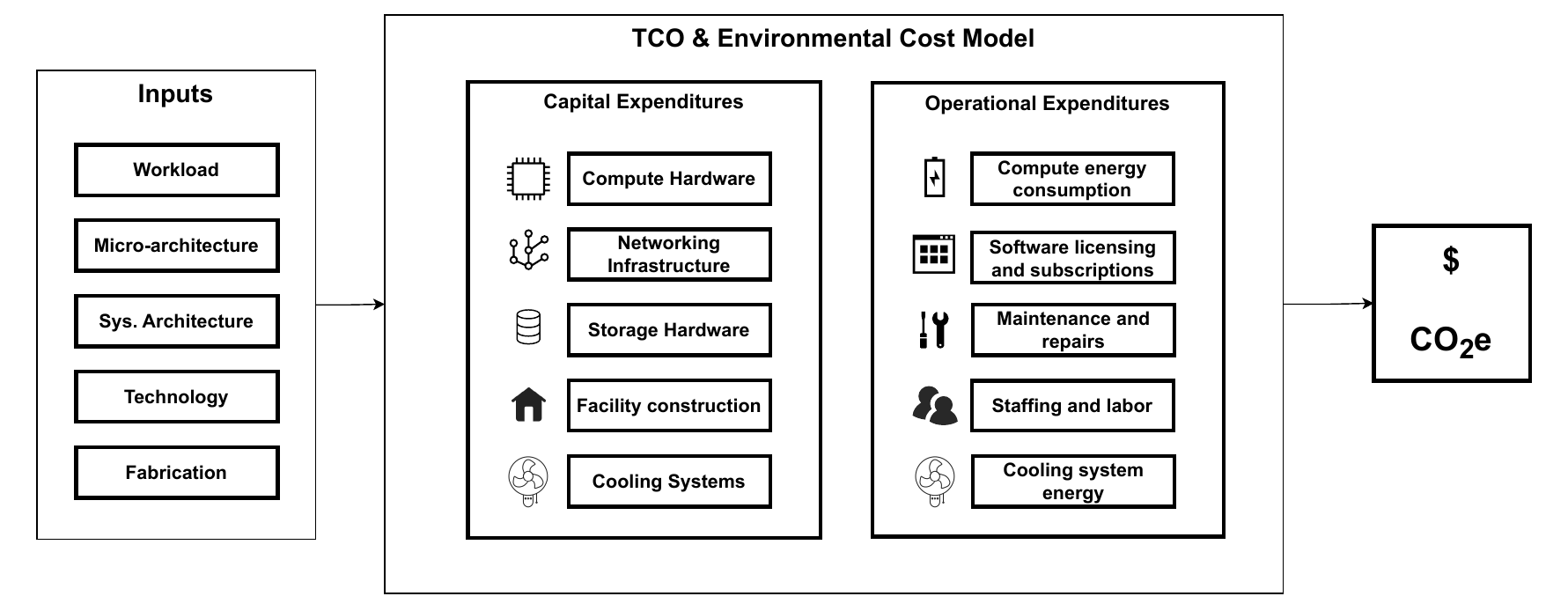}
    \caption{Summary of a typical TCO modeling flow. The \emph{Inputs} box covers the compute stack from workload to fabrication. The \emph{TCO and Environmental Cost} box lists common CapEx/OpEx components, with outputs in currency or CO$_2$ equivalent emissions (CO$_2$e).}
    \label{fig:tco_summary}
\end{figure*}

TCO is typically defined as the sum of \emph{Capital Expenditure} (CapEx) and \emph{Operating Expenditure} (OpEx) over a system’s lifetime~\cite{barroso2019datacenter} (see Figure~\ref{fig:tco_summary}). CapEx covers upfront costs for compute hardware and datacenter infrastructure, including buildings, power delivery, networking, and cooling systems—costs that vary by size, location, and workload. OpEx includes ongoing operational costs, with energy typically the dominant factor~\cite{schneider2025life}. This depends on component power use and datacenter Power Usage Effectiveness (PUE), the ratio of total facility power to IT power. While some hyper-scale centers report PUE as low as 1.1~\cite{googledatacenter}, the 2024 global average is 1.56~\cite{uptimeinstitutesurvey2024}. Cooling overhead is commonly estimated using server Thermal Design Power (TDP), and energy costs are derived from local electricity prices and energy sources. Other non-computing-related expenses, such as staffing, system maintenance, and software licensing, are also part of OpEx, but are beyond the scope of this survey.

\subsection{Environmental Cost}
Alongside financial costs, the environmental impact of DNN training systems has become an increasingly important metric, particularly in light of the climate crisis and sustainability goals. This concern is especially acute for LLMs,  which carry substantial environmental costs due to the high energy consumption and significant greenhouse gas (GHG) emissions associated with hyperscale data centers~\cite{schneider2025life, luccioni2023estimating, wu2024beyond, wu2022sustainable}.

Similarly to TCO, environmental costs can be divided into \emph{Embodied Emissions} (CapEm) and \emph{Operating Emissions} (OpEm). The term CapEm is used here for symmetry with TCO terminology. CapEm is connected to the embodied carbon emissions from manufacturing and recycling~\cite{gupta2021chasing} and OpEm is dependent on the underlying technologies and the complexity of the system configuration~\cite{patterson2021carbon}. Environmental costs are typically captured using emissions models, which estimate the carbon footprint associated with both the manufacturing and operation of computing systems.

\subsubsection{Carbon Emissions Metrics}
GHG emissions are typically reported in grams of carbon dioxide equivalent (CO$_2$e), which standardizes the impact of different gases based on their global warming potential. To facilitate comparison across hardware platforms or workloads, the \emph{Compute Carbon Intensity} (CCI) metric has been proposed. CCI normalizes emissions by the amount of computation performed, and can, for example, be expressed in emissions per floating-point operation (CO$_2$e/FLOP). This enables more meaningful comparisons of hardware efficiency and environmental impact, as demonstrated in Google's life-cycle assessment of Tensor Processing Unit (TPU) generations~\cite{schneider2025life}.

Another relevant metric is the \emph{Total Carbon Delay Product} (tCDP). This metric captures the joint impact of CO$_2$e emissions across the lifetime of a product and the execution time of a task. It penalizes both high carbon emissions and slow execution. It is especially relevant when optimizing for both environmental and performance efficiency, as shown in~\cite{elgamal2025cordoba}.

\subsubsection{Geographic Variability}

Geographic location significantly affects both TCO and environmental impact. Emissions for large distributed systems can vary up to 10× depending on deployment region~\cite{patterson2021carbon}. Schneider et al.~\cite{schneider2025life} show this effect in a life-cycle analysis of Google TPUs, where access to carbon-free energy (CFE) substantially reduced operational emissions.

While this survey focuses on compute-related factors, geographic variability remains important. Standardizing energy (kWh) and emissions (CO$_2$e) reporting can help decouple location effects and enable more comparable system-level analyses.

\subsection{Comparison of TCO and Environmental Costs Models}

The space of TCO and environmental costs models is highly diverse, with each model targeting different aspects of system cost or environmental impact. To support navigation of this landscape, a comprehensive table of available models is presented (see Table~\ref{tab:tco_emissions_summary}), offering a high-level overview of their capabilities. Each model is categorized by its primary focus—TCO or emissions—and by whether it supports modeling of CapEx, OpEx, CapEm, and OpEm. Beyond these fundamentals, the models are further classified based on their treatment of six key modeling dimensions:

\begin{itemize}
\item \textbf{Fabrication} – Level of semiconductor manufacturing modeling, including resource usage (e.g., water, energy, and raw materials), process node, and defect density. This could also include further manufacturing steps, such as packaging and assembly. 
\item \textbf{Technology} – Modeling of hardware-level features, such as die area, die stacking, or on-chip interconnect, which affect system behavior and cost.
\item \textbf{Network} – Type of internode communication modeling, including representation of interconnects, bandwidth, and switching infrastructure.
\item \textbf{Storage} – Approach to modeling network-connected persistent storage configurations and their impact on cost and emissions.
\item \textbf{Node Architecture} – Method used to model compute node characteristics, including specification-based inputs (e.g., datasheet values), analytical models, or detailed microarchitectural simulation.
\item \textbf{Workload} – Type of workload used in TCO and emissions model.
\end{itemize}

While some models also account for broader economic factors such as facility infrastructure, staffing, non-recurring engineering (NRE), and R\&D costs, these dimensions are excluded from the table to maintain focus on the compute stack.

\afterpage{
\begin{table*}[htbp]
\caption{Comparison of recent models for Total Cost of Ownership (TCO) and carbon emissions in computing systems. The table summarizes licensing, data availability, modeling of capital and operational expenditures and emissions, and coverage of key system components such as fabrication processes, technology characteristics, and workload assumptions.}
\ra{1.2}
\tiny
\begin{tabular}{
|>{\centering\arraybackslash}p{0.25cm}|p{1.6cm}p{1.2cm}C{0.5cm}|
C{0.5cm}C{0.5cm}|C{0.5cm}C{0.5cm}|
>{\raggedright}p{2.2cm}>{\raggedright}p{1.7cm}>{\raggedright}p{1.3cm}>{\raggedright}p{1.2cm}>{\raggedright}p{2cm}>{\raggedright\arraybackslash}p{2.2cm}|}
\hline
\yycell{2}{\textbf{Year}} & \multicolumn{3}{c|}{\textbf{MODEL INFO}}& \multicolumn{2}{c|}{\textbf{TCO}} & \multicolumn{2}{c|}{\textbf{EMISSIONS}} & \multicolumn{6}{c|}{\textbf{MODEL COMPONENTS}}  \\ 
\hhline{~|---|--|--|------}
& \textbf{Name} & \textbf{License} & \textbf{Data Avail.} & \textbf{CapEx} & \textbf{OpEx} & \textbf{CapEm} & \textbf{OpEm}  & \textbf{Fabrication} & \textbf{Technology} & \textbf{Network} & \textbf{\mr{Storage}} & \textbf{Node Architecture} & \textbf{Workload} \\ 
\hline
\yycell{1}{\textquotesingle13} & EETCO~\cite{hardy2013analyticaleetco} & Closed Source & \bad & \good & \good & \bad & \good &\nota&Spec-based&Spec-based& Spec-based &Spec-based&\nota\\ 
\hdashline[1pt/1pt]
\yycell{2}{\hspace{-2ex}{2017}}& Baidu~\cite{cui2017total} & Closed Source & \bad & \good & \good & \bad & \bad &\nota&Spec-based&Spec-based&\nota&Spec-based&\nota\\ 
& Moonwalk~\cite{khazraee2017moonwalk} & Closed Source & \bad & \good & \good & \bad & \bad & Foundry process steps, packaging, assembly & Die area scaling &\nota&\nota& Custom analytical model & Crypto mining, Transcoding, DNN \\
\hdashline[1pt/1pt]
\yycell{2}{2019} & \href{https://github.com/Pitt-JonesLab/Greenchip}{GreenChip}~\cite{kline2019greenchip} & BSD-3 & \good & \bad & \bad & \good & \good & LCA-based & Die area scaling &\nota&\nota& Sniper, McPAT, CACTI & SPEC, NAS \\
& \href{https://github.com/mlco2/impact}{MLCO2}~\cite{lacoste2019quantifying} & MIT & \good & \bad & \bad & \bad & \good &\nota&Spec-based&\nota&\nota&Spec-based& DNN \\
\hdashline[1pt/1pt]
\yycell{1}{\textquotesingle20} & \href{https://github.com/Breakend/experiment-impact-tracker}{Henderson et al.}~\cite{henderson2020towards} & MIT & \good & \bad & \bad & \bad & \good &\nota&\nota&\nota&\nota& Hardware profiling & DNN \\
\hdashline[1pt/1pt]
\yycell{2}{\hspace{-5ex}{2022}} & \href{https://github.com/alugupta/ACT}{ACT}~\cite{gupta2022act} & MIT & \good & \bad & \bad & \good & \good & Foundry process steps, packaging & Die area scaling &\nota& \nota & Hardware profiling,Spec-based& Mobile, DNN \\
& \href{https://netzero.imec-int.com/}{imec.netzero}~\cite{imecnetzero} & Closed Source & \bad & \good & \bad & \good & \good & Foundry process steps, packaging & Die area scaling &\nota& \nota &Spec-based& Mobile, HPC \\
\hdashline[1pt/1pt]
\yycell{3}{\hspace{-2ex}{2023}} & \href{https://github.com/facebookresearch/CarbonExplorer}{Carbon Explorer}~\cite{acun2023carbon} & CC BY-NC 4.0 & \good & \bad & \bad & \good & \good & LCA-based &Spec-based&\nota& \nota &Spec-based&\nota\\
& \href{https://github.com/bigscience-workshop/carbon-footprint}{Luccioni et al.}~\cite{luccioni2023estimating} & No license & \good & \bad & \bad & \good & \good & LCA-based &\nota&\nota&\nota&Spec-based& LLM \\
& \href{https://github.com/PrincetonUniversity/ttm-cas}{Ning et al.}~\cite{ning2023supply} & BSD-3 & \good & \bad & \bad & \bad & \bad & Foundry throughput, packaging, assembly, testing & Die area scaling &\nota&\nota&Spec-based&\nota\\
\hdashline[1pt/1pt]
\yycell{6}{\hspace{-10ex}{2024}} & Chiplet Cloud~\cite{peng2024chipletcloudbuildingai} & Closed Source & \bad & \good & \good & \bad & \bad & Spec-based & Die area scaling &\nota&\nota& Custom analytical model & LLM \\
& \href{https://github.com/mlco2/codecarbon}{CodeCarbon} & MIT & \good & \bad & \bad & \bad & \good &\nota&\nota&\nota&\nota& Hardware profiling & DNN \\
& \href{https://zenodo.org/records/10896255}{GreenSKU}~\cite{wang2024designinggreensku} & CC BY 4.0 & \good & \bad & \bad & \good & \good & imec.netzero, Makersite & Die area scaling, Spec-based & Spec-based & Spec-based &Spec-based& Redis, web, inference, speech \\
& \href{https://github.com/SotaroKaneda/MLCarbon}{LLMCarbon}~\cite{faiz2024llmcarbon} & Apache 2.0 & \good & \bad & \bad & \good & \good &  CO$_2$e per area & Die area scaling &\nota&\nota& Analytical flop \& efficiency model & LLM training \\
& \href{https://github.com/answers111/OpenCarbonEval}{OpenCarbonEval}~\cite{yu2024opencarboneval} & No license & \good & \bad & \bad & \good & \good & CO$_2$e per area &Die area scaling&\nota&\nota&Analytical flop \& dynamic throughput model& Vision models, LLM training \\
& \href{https://github.com/arc-research-lab/SCARIF}{SCARIF}~\cite{ji2024scarif} & No license & \good & \bad & \bad & \good & \good & ACT-, GreenChip-based & Die area scaling, Spec-based &\nota& \nota &Spec-based& DNN inference \\
\hdashline[1pt/1pt]
\yycell{4}{\hspace{-5ex}{2025}} & \href{https://github.com/nanocad-lab/CATCH}{CATCH}~\cite{graening2025catch} & Apache 2.0 & \good & \good & \good & \bad & \bad & Foundry process steps, packaging, assembly, testing & Die area scaling, 2.5D/3D stacking, IO &\nota&\nota& Custom chiplet stacking &\nota\\
& \href{https://github.com/facebookresearch/CATransformers}{CATransformers}~\cite{wang2025carbon} & CC BY-NC 4.0 & \good & \bad & \bad & \good & \good & ACT-based & Die area scaling &\nota&\nota& SCALE-sim & Multimodal inference \\
& CORDOBA~\cite{elgamal2025cordoba} & Closed Source & \bad & \bad & \bad & \good & \good & ACT-based & Die area scaling &\nota&\nota& Analytical simulator & AI, XR kernels \\
& GreenScale~\cite{kim2023greenscale} & Closed Source & \bad & \bad & \bad & \good & \good & ACT-, LCA-based & ACT-, LCA-based &LCA-, spec-based&\nota&Spec-based& AR/VR, Games, Mobile DNN \\
\hline
\end{tabular}
\label{tab:tco_emissions_summary}
\par
\end{table*}
}

\subsection{Takeaways and Future Opportunities}

\begin{takeaway}
Limited access to high-quality embodied and operational cost data remains a key obstacle.
\end{takeaway}

The growing scale and environmental impact of LLMs has increased interest in TCO and emissions modeling for distributed DNN systems. Industry disclosures~\cite{patterson2021carbon, wu2022sustainable, faiz2024llmcarbon}, in-depth analyses~\cite{gupta2021chasing}, and tools like ACT~\cite{gupta2022act} have advanced the field, supported by proposals for unified energy reporting frameworks~\cite{henderson2022systematic}.

However, limited data quality and availability—especially for embodied emissions tied to chip manufacturing and infrastructure—remains a key barrier. Proprietary cost structures and fabrication processes hinder accurate modeling. Even high-profile estimates, such as Strubell et al.’s CO$_2$e numbers~\cite{strubell2019}, have been revised significantly when more detailed data became available~\cite{patterson2021carbon}.

Cost modeling at the device level also suffers from opaque CapEx figures and variable operational parameters. Cloud providers suggest better long-term TCO than on-premises setups~\cite{azuretco}, though generalization is difficult. Recent transparency efforts, including data disclosures by hyperscalers and analyses from imec~\cite{bardon2020dtco, imecnetzero}, mark promising steps forward.

A related challenge lies in device-level cost modeling. The gap between retail pricing and actual capital expenditure for custom silicon is significant, while key operational parameters—such as energy consumption and device lifetime—are often undisclosed and deployment-specific. For instance, TCO estimates from providers like AWS~\cite{azuretco} indicate that cloud-based deployments can achieve greater long-term cost-efficiency than on-premises systems. Efforts toward transparency—such as infrastructure data disclosures by hyperscalers and analyses from organizations like imec, including studies of emissions across technology nodes~\cite{bardon2020dtco} and the imec.netzero platform~\cite{imecnetzero}—represent meaningful progress.

The lack of data also significantly hinders the validation of TCO and emissions models, which is challenging without access to detailed industry-level information.

\begin{takeaway}
Specification and datasheet based TCO and emissions models work for current systems but don't scale to future technologies and ML models.
\end{takeaway}

Many TCO and emissions models do not use any fabrication parameters and instead focus on characterizing existing hardware. This includes using off-the-shelf pricing~\cite{hardy2013analyticaleetco,cui2017total}, referencing publicly available industry reports or hardware specifications ~\cite{kline2019greenchip,luccioni2023estimating,acun2023carbon,ji2024scarif,peng2024chipletcloudbuildingai}, or extracting power consumption by hardware profiling~\cite{lacoste2019quantifying,henderson2020towards,benoit_courty_2024_11171501}. This works well for estimating existing systems, but does not scale to future process technologies, hardware architectures, or ML models. 

The performance, cost, and emissions of systems based on future technologies can be estimated by scaling state-of-the-art process nodes~\cite{khazraee2017moonwalk,ning2023supply,peng2024chipletcloudbuildingai}, or by modeling process-level manufacturing parameters~\cite{imecnetzero,gupta2022act,graening2025catch}. However, although process-level modeling may be more accurate, it relies on sensitive manufacturing information which is not readily available.

\begin{takeaway}
    TCO and emissions are rarely modeled together, limiting the ability to capture trade-offs between cost, efficiency, and environmental impact.
\end{takeaway}

DNN system research has traditionally prioritized performance and accuracy. As system scale has increased, both TCO and emissions have become critical metrics. Yet, most TCO studies overlook emissions, and emissions-focused work rarely models TCO—likely due to differing community priorities or the relative opacity of cost data compared to emissions disclosures.

Early TCO studies, before the AI era, focused on financial modeling for moderately sized data centers~\cite{koomey2008,hardy2013analyticaleetco,cui2017totalcost}. With the rise of AI, hyper-scale data centers have grown, drawing attention to their resource consumption~\cite{luccioni2023estimating}. Operators can estimate OpEm using internal energy data~\cite{patterson2021carbon,wu2022sustainable}, and reporting OpEm alongside OpEx would be valuable. Recent efforts to cut emissions go beyond efficiency, exploring carbon-aware scheduling~\cite{acun2023carbon,kim2023greenscale} and low-carbon server designs~\cite{wang2024designinggreensku}.

Jointly modeling TCO and emissions could incentivize the transition to systems that emit less CO$_2$e, by making environmental impact a first-class metric in system design and evaluation.

\begin{takeaway}
Current TCO and carbon models often overlook the impact of scheduling, reliability, and workload variation over a hardware's lifetime.
\end{takeaway}

Existing work has established that workload characteristics significantly influence hardware utilization, and consequently, overall operational expenditures and embodied emissions. Models such as OpenCarbonEval~\cite{yu2024opencarboneval}, LLMCarbon~\cite{faiz2024llmcarbon}, Chiplet Cloud~\cite{peng2024chipletcloudbuildingai}, and CATransformers~\cite{wang2025carbon} incorporate workload representations into their estimations. However, variations in scheduling policies and runtime orchestration can also substantially impact workload performance~\cite{wongpanich2025machine}. Additionally, reliability issues and unforeseen interruptions mean that effective training time often deviates from the ideal~\cite{grattafiori2024llama3herdmodels}. Furthermore, over the lifetime of a piece of hardware, multiple workloads may be run—interleaved with periods of downtime—which could also be accounted for.
These factors highlight a notable gap in current TCO and carbon modeling research. While many of these aspects could be integrated into existing models, doing so often requires access to production-quality data, which remains a key limitation.

\begin{takeaway}
The lack of integration between TCO and emissions models and dynamic distributed DNN simulators limits their usefulness for evaluating future technologies and system designs.
\end{takeaway}

Integrating TCO and environmental models with distributed DNN training simulators is key to enabling forward-looking research. This allows more accurate estimation of future workload and hardware impacts, as component-level analysis alone often misses full system costs~\cite{patterson2021carbon}.

Some recent distributed DNN training simulators, such as Calculon~\cite{calculon}, DeepFlow~\cite{ardalani2024deepflow}, LLMCompass~\cite{llmcompass}, and Optimus~\cite{optimus}, have begun to incorporate broader cost modeling (see Table~\ref{tab:dnnsim}), though none currently account for environmental impact. Conversely, emerging TCO and emissions models such as CATransformers~\cite{wang2025carbon} and CORDOBA~\cite{elgamal2025cordoba} include workload and microarchitectural inputs, but remain decoupled from dynamic simulation infrastructure (see Table~\ref{tab:tco_emissions_summary}). LLMCarbon~\cite{faiz2024llmcarbon} and Chiplet Cloud~\cite{peng2024chipletcloudbuildingai} demonstrate end-to-end modeling from workload to TCO/emissions, but rely on coarse-grained workload and performance models. Integrating such models with operator-level distributed DNN simulators with more detailed analytical and profiling-based approaches could improve their applicability to future hardware and system evaluations.

Bridging this gap presents a significant opportunity for future research, though it spans multiple abstraction levels and requires cross-disciplinary expertise. As a practical starting point, existing distributed DNN simulators could be integrated with available TCO and emissions models to enable more holistic system-level evaluation.
\section{Conclusions and Future Work} \label{sec:takeaways}

\ifthenelse{\boolean{is_acm}}{

\begin{table*}[htbp]
\centering
\tablefont

\begin{tabular}{@{}>{\raggedright\arraybackslash}p{2.5cm}@{\hspace{1em}} >{\raggedright\arraybackslash}p{5.5cm}@{\hspace{1em}} >{\raggedright\arraybackslash}p{5.5cm}@{}}
\toprule
\textbf{Domain} & \textbf{Technical Trends} & \textbf{Opportunities} \\
\midrule

\multirow{3}{*}{\parbox[c][\dimexpr 5\baselineskip][c]{3.5cm}{\vspace{3.5\baselineskip}\raggedright Workload\\Representation}}

& Shift from configuration-based to operator-level IRs for better fidelity. 
& Extend IRs to cover system-level events such as I/O and checkpointing. \\ \addlinespace
& Custom IRs dominate due to simplicity and simulator-specific needs. 
& Develop modular, reusable IRs to reduce fragmentation across simulators. \\ \addlinespace
& MLIR emerging as a common IR across compiler stacks, but currently underutilized in distributed DNN simulators.
& Explore MLIR-based representations for distributed simulation workloads. \\ \addlinespace

\midrule

\multirow{4}{*}{\parbox[c][\dimexpr 5\baselineskip][c]{3.5cm}{\vspace{4.7\baselineskip}\raggedright Distributed DNN\\Training Simulators}}

& Profiling-based simulation gaining traction for fidelity-speed tradeoff. 
& Improve scalability and trace collection to support large model profiling. \\ \addlinespace
& Simulator validation remains limited, especially at scale. 
& Develop methodologies and benchmarks for cross-platform simulator validation. \\ \addlinespace
& Focus on LLMs dominates simulator development. 
& Generalize simulation frameworks to support diverse DNN architectures and workloads. \\ \addlinespace
& Memory models are mostly implicit and network models analytical. 
& Incorporate detailed network and memory models in existing simulators. \\ \addlinespace

\midrule

\multirow{5}{*}{\parbox[c][\dimexpr 5\baselineskip][c]{3.5cm}{\vspace{6.9\baselineskip}\raggedright TCO \&\\Emissions
}}
& CapEx and CapEm data remains difficult to access and model. 
& Build community datasets or synthetic modeling tools for missing cost/emissions data. \\ \addlinespace
& Most models are backward-looking or specification-based. 
& Develop predictive TCO/emissions models for emerging technologies and workloads. \\ \addlinespace
& Workload-specific effects on OpEx/OpEm are underexplored. 
& Incorporate lifecycle-aware workload modeling into TCO/emissions analysis. \\ \addlinespace
& Distibuted DNN simulators and TCO/emissions models are not considered together. 
& Enable end-to-end flows coupling distributed DNN training simulators with energy/cost models for design-space exploration. \\

\bottomrule
\end{tabular}
\end{table*}

} {

\begin{itemize}
  \item \textbf{Workload Representation:} There is a clear shift from configuration-based to operator-level IRs to improve modeling fidelity. However, current IRs are mostly simulator-specific and lack system-level event coverage. Opportunities include developing modular, reusable IRs and extending them to capture system-level behaviors like I/O and checkpointing. While MLIR is gaining adoption across compiler stacks, it remains underutilized in distributed DNN simulation—exploring MLIR-based IRs could improve interoperability.

  \item \textbf{Distributed DNN Training Simulators:} Profiling-based simulation is increasingly used to balance fidelity and speed, but collecting large-scale traces and ensuring simulator scalability remains challenging. Simulator validation, especially at scale, is still limited. There's also a strong focus on LLMs, which narrows generalizability. Future work should address simulator validation, extend support to diverse DNN workloads, and integrate detailed memory and network models.

  \item \textbf{TCO and Emissions Modeling:} Data for CapEx and CapEm remains scarce and difficult to model. Most current tools are retrospective or rely on static specifications. There is a need for predictive models that address emerging technologies and workloads, as well as more accurate modeling of workload-specific impacts on OpEx and OpEm. Finally, distributed DNN training simulators and TCO/emissions models are typically developed in isolation—future work should focus on integrating these to enable holistic, end-to-end design-space exploration.
\end{itemize}

}

\section*{Acknowledgments}
This work is funded by the Advanced Research + Invention Agency (ARIA).  
The authors would like to acknowledge the following persons for the contributions and discussions: Matthew Walker, Sahan Gamage, Vinay Kumar Baapanapalli Yadaiah and Arindam Mallik. 
The authors used GPT-4o for language editing and proofreading parts of the manuscript.

\bibliographystyle{IEEEtran}
\bibliography{references}

\end{document}